# Shocked $H_2$ and $Fe^+$ dynamics in the Orion bullets

Jonathan A. Tedds,[1]★† Peter W. J. L. Brand[1] and Michael G. Burton[2]
[1]*Institute for Astronomy, University of Edinburgh, Royal Observatory, Blackford Hill, Edinburgh EH9 3HJ*
[2]*School of Physics, University of New South Wales, Sydney, NSW 2052, Australia*

astro-ph/9903022 v2   26 Sep 1999

**ABSTRACT**
Observations of $H_2$ velocity profiles in the two most clearly defined Orion bullets are extremely difficult to reconcile with existing steady-state shock models. We have observed [Fe II] 1.644-μm velocity profiles of selected bullets, and $H_2$ 1–0 S(1) 2.122-μm velocity profiles for a series of positions along and across the corresponding bow-shaped shock fronts driven into the surrounding molecular cloud. Integrated [Fe II] velocity profiles of the brightest bullets are consistent with theoretical bow shock predictions. However, observations of broad, singly peaked $H_2$ 1–0 S(1) profiles at a range of positions within the most clearly resolved bullet wakes are not consistent with molecular shock modelling. A uniform, collisionally broadened background component which pervades the region in both tracers is inconsistent with fluorescence arising from the ionizing radiation of the Trapezium stars alone.

## 1 INTRODUCTION

The nature of molecular shocks, which play an important role in the processes of momentum and energy transfer within starforming molecular clouds (McKee 1989), is still uncertain (Draine & McKee 1993). The Orion molecular cloud is the brightest known source of shocked $H_2$ emission, and as such has been the primary testbed for theoretical models. In this paper we describe new observations of shocked $H_2$ in Orion, and analyse the fundamental obstacle to modelling its excitation: namely the extremely broad velocity profiles observed. With the discovery of the Orion bullets and associated $H_2$ wakes, we describe an experiment to test previous models to breaking point by detailed observations of $Fe^+$ and $H_2$ line profiles at high spectral and spatial resolution using CGS4 with the new, larger and more sensitive 256×256 pixel array at the United Kingdom InfraRed Telescope (UKIRT).

The outflow associated with the Kleinmann–Low (KL) infrared nebula in Orion is the best studied, as it is close (480 pc) and bright, permitting higher spatial resolution and signal-to-noise ratios than are possible in other massive star-forming regions. Genzel & Stutzki (1989) reviewed the parameters of the outflowing gas, and distinguished a low-velocity (18 km s$^{-1}$) 'expanding doughnut' flow, a high-velocity flow or 'plateau' and high-velocity shocked gas. Each component is identified with a different pattern of spatial distribution, kinematics and excitation. Following the initial discovery, Beckwith et al. (1978) mapped the $H_2$ 1–0 S(1) line emission over a 2 × 2 arcmin$^2$ field about Becklin–Neugebauer (BN)–KL at 5-arcsec spatial resolution, identifying the brightest region as Peak 1 (OMC-1) centred on $05^h32^m46^s.09$, $-05°23'57".13$ (1950). Measurements of the intensities of the brightest near-infrared $H_2$ emission lines at OMC-1 (Beckwith et al. 1978) gave a vibrational $H_2$ excitation temperature of $T_{vib} = 2000 \pm 300$ K for a thermalized Boltzmann distribution of level populations with an average column density of $\sim 10^{19}$ cm$^{-2}$. Pure radiative excitation, as described by Black & Dalgarno (1976), was ruled out.

Models for the $H_2$ excitation were advanced involving planar hydrodynamic jump (J) shocks (Hollenbach & Shull 1977; Kwan 1977; London, McCray & Chu 1977). These models give shocked $H_2$ line ratio predictions including that the 1–0 S(1)/2–1 S(1) line ratio is of the order of ~10 for a 10 km s$^{-1}$ shock wave moving into gas of density 10$^5$ cm$^{-3}$. However, such a shock must dissociate $H_2$ at velocities exceeding $\sim 24$ km s$^{-1}$ (Kwan 1977). Therefore, when Nadeau & Geballe (1979) observed individual $H_2$ line profiles in this region having full width at zero intensity (FWZI) linewidths exceeding 150 km s$^{-1}$, this model was quickly brought into doubt. Since molecular clouds have measurable magnetic fields carried by ions, such difficulties might be overcome by modelling magnetohydrodynamic shocks having continuous C-type shock fronts (Draine 1980; Draine & Roberge 1982; Chernoff, McKee & Hollenbach 1982; Draine, Roberge & Dalgarno 1983).

★ E-mail: jat@ast.leeds.ac.uk
† Present address: Department of Physics and Astronomy, University of Leeds, Woodhouse Lane, Leeds LS2 9JT.

A major problem of fitting planar C-shocks to the Orion observations is that the predicted excitation spectrum greatly underestimates the higher energy level populations (and hence column densities) observed towards OMC-1 (Brand et al. 1988). In addition, such a model predicts that the line ratios of two high-excitation lines must vary considerably given small changes in local physical conditions, whereas Brand et al. (1989a) demonstrated that the ratio of the 1–0 O(7) and 0–0 S(13) $H_2$ lines, at excitation energies of 8300 and 17 000 K respectively, remains constant over the outflow in a 5-arcsec beam. These two difficulties are naturally explained by a simple J-shock (Brand et al. 1988) assuming higher density gas than the early models. Of course, such a model is inconsistent with the observed profiles and also implies that $H_2$ is the dominant coolant compared with $H_2O$ or CO.

To reconcile this contradictory evidence, it was suggested that several shocks were being observed along a given line of sight. If the outflow is at least partly composed of dense clumps of material either ejected or swept up by a wind, then a bow shock will form around the leading edge of each clump as it drives through the molecular gas. The shock strength and hence degree of excitation vary with the normal component of the clump velocity, depending on the geometry of the bow. Hence, in an observation that does not spatially resolve individual clumps, the complete range of excitation conditions is observed, from a dissociated cap with atomic line emission, through to molecular line emission from cooler gas further down the wake. Line ratios would also be constant if all bows conformed to a generic shape. A series of papers (Smith & Brand 1990a,b,c; Smith, Brand & Moorhouse 1991) successfully accounted for a number of observational features with such bow C-shock models.

However, the fast bow C-shock solutions create their own difficulties. To fit the broad-FWZI $H_2$ 1–0 S(1) profile width of 140 km s$^{-1}$ observed with a 5-arcsec aperture at OMC-1 by Brand et al. (1989b) requires C-shocks that do not dissociate $H_2$ at velocities in excess of 60–70 km s$^{-1}$, and therefore that the bow geometry is assumed rather than self-consistently determined. This, in turn, implies a very high magnetic field of $\gtrsim 50$ mG, higher than recent estimates of $\sim 10$ mG determined from the dispersion of the position angles of the $H_2$ polarization vectors (Chrysostomou et al. 1994) and implying very high magnetic pressures. Also, the full range of observed shocked $H_2$ excitation conditions must be sampled within the 5-arcsec beamsize (Brand et al. 1988), placing tight constraints on the (unknown) bow geometry and size at OMC-1.

Recent near-infrared imaging of Orion with 0.5-arcsec spatial resolution in the emission lines of $H_2$ (2.122 $\mu$m) and [Fe II] (1.644 $\mu$m) has significantly advanced our view of the shocked molecular outflow (Allen & Burton 1993). Many new Herbig–Haro-type objects were revealed, visible as [Fe II] bullets, at the heads of wakes of $H_2$-emitting gas. The bullets (apparently originating within 5 arcsec of IRc2) have been ejected over a wide opening angle. Considering also the [O I] 6300 Å linewidths of up to 380 km s$^{-1}$ measured towards the brightest bullets previously identified in this region (Axon & Taylor 1984), an explosive origin in the core of Orion within the last $10^3$ yr was suggested by Allen & Burton (1993).

Crucially for molecular shock studies, the newly discovered bullets and wakes are resolved spatially to be typically 2–4 arcsec in size and the wakes or 'fingers' a factor of 2 or so longer behind them. This corresponds to a scale of 0.005 to 0.01 pc for each bullet at the distance of Orion. The widths of the shock fronts remain unresolved. However, given the 1.2-arcsec spatial and 23 km s$^{-1}$ FWHM spectral resolution achieved using CGS4 with the echelle and 256×256 pixel array, it has become possible to search for the dynamical variations expected within the Orion bullet $H_2$ wakes, if they are indeed bow shocks.

Given the known physical constraints on conditions in Orion, we expect to see the $H_2$ shock velocity vary from the maximum for J- or C-shocks ($25 \lesssim v_s \lesssim 50$ km s$^{-1}$) to the minimum necessary to excite the lines at all ($\gtrsim 8$ km s$^{-1}$). Each line of sight will sample a range of conditions (and velocities), and so this must be incorporated into any interpretation. This problem has been carefully analysed for atomic line emission by Hartigan, Raymond & Hartmann (1987), who showed that the FWZI of the integrated profile over a single bow shock must be equal to the speed of the bullet itself. By also measuring the maximum and minimum velocities, one can further deduce the orientation to the line of sight. Therefore measurement of [Fe II] profiles for a bullet provides an estimate of the velocity and orientation of the associated $H_2$ bow, and it is then possible to compare measured and expected $H_2$ shock velocities as they vary with position in the bow. In this paper, we present measurements of the [Fe II] and $H_2$ velocity profiles in the two most clearly defined Orion bullets. We will present measurements of the associated $H_2$ excitation conditions in both of these wakes in a forthcoming paper.

## 2 OBJECT NOMENCLATURE

In this paper we refer to the bullets and associated wakes in Orion by a system of designation of compact sources and stars in M42 based on the J2000 position on the sky, as proposed by O'Dell & Wen Zheng (1994). This is necessary as there is now a proliferation of observations in this region, many having separate classification schemes. The first three digits indicate the position of right ascension (J2000), and the second three digits indicate the declination, with the box so designated subtending 1.5 arcsec in right ascension and 1.0 arcsec in declination. The common values for the inner region of M42 of $5^h35^m$ and $-5°20'$ are not included. Hence the bullet located at $5^h35^m12\rlap{.}^s6$, $-5°20'53''$ is designated M42 HH126–053, and the bullet at $5^h35^m12\rlap{.}^s0$, $-5°21'14''$ is designated M42 HH120–114. Absolute slit positions within each of the bullets and wakes are listed in Table 1, and the locations of the slits are displayed in Figs. 1 and 2, superimposed on the narrow-band $H_2$ 1–0 S(1) and [Fe II] 1.644-$\mu$m images from Allen & Burton (1993).

## 3 OBSERVATIONS AND DATA REDUCTION

High-resolution, near-infrared, long-slit echelle spectra were measured of the [Fe II] $a^4F_{5/2}$–$a^4D_{7/2}$ line profile at 1.644 $\mu$m on two of the most prominent bullets associated with the most clearly defined $H_2$ wakes. $H_2$ 1–0 S(1) profiles were then measured at a range of positions along and across the associated wakes. The observations of the [Fe II] emission from the bullet M42 HH126–053 (previously denoted M42 HH7) were carried out at UKIRT on the night of 1992 September 19 with the cooled grating spectrometer CGS4 and the original (62×58 pixel) InSb detector array (Mountain et al. 1990). The [Fe II] observations for the bullet M42 HH120–114 (no previous designation) and the observations of the $H_2$ wakes associated with both bullets were carried out at UKIRT on the nights of 1995 October 5–8 with CGS4 and a

**Table 1.** Absolute position (1950) on the sky of row 25 for each slit position within each bullet wake. The pixel scale is expressed as length × slit width.

| Target | Tracer | Slit label | Absolute position of slit row 25 (1950) | | Slit orientation /° W of N | Pixel scale /arcsec | Sky offset /arcsec west |
|---|---|---|---|---|---|---|---|
| | | | RA | Dec. | | | |
| M42 HH126–053 | $Fe^+$ | A | $05^h 32^m 44.58^s$ | $-05° 22' 32.5''$ | +25.5 | 2.16×1.1 | 300 |
| | $H_2$ | B | $05^h 32^m 45.52^s$ | $-05° 22' 52.7''$ | +25.5 | 1.74×0.9 | 600 |
| | | C | $05^h 32^m 45.32^s$ | $-05° 22' 52.7''$ | +25.5 | 1.74×0.9 | 600 |
| | | D | $05^h 32^m 45.18^s$ | $-05° 22' 52.7''$ | +25.5 | 1.74×0.9 | 600 |
| | | E | $05^h 32^m 45.05^s$ | $-05° 22' 52.7''$ | +25.5 | 1.74×0.9 | 600 |
| M42 HH120–114 | $Fe^+$ | F | $05^h 32^m 44.37^s$ | $-05° 23' 06.7''$ | +22.0 | 1.69×0.9 | 600 |
| | | G | $05^h 32^m 44.30^s$ | $-05° 23' 06.7''$ | +22.0 | 1.69×0.9 | 600 |
| | $H_2$ | H | $05^h 32^m 44.86^s$ | $-05° 23' 10.5''$ | +41.0 | 1.74×0.9 | 600 |
| | | I | $05^h 32^m 44.75^s$ | $-05° 23' 11.9''$ | +41.0 | 1.74×0.9 | 600 |
| | | J | $05^h 32^m 44.64^s$ | $-05° 23' 13.4''$ | +41.0 | 1.74×0.9 | 600 |
| | | K | $05^h 32^m 44.53^s$ | $-05° 23' 14.8''$ | +41.0 | 1.74×0.9 | 600 |
| | | L | $05^h 32^m 44.42^s$ | $-05° 23' 16.3''$ | +41.0 | 1.74×0.9 | 600 |

256×256 pixel array, used in echelle mode with the short 150-mm focal length camera, giving a pixel size of ∼1.7 arcsec (exact scale depending on wavelength) along the slit and ∼0.9 arcsec in the dispersion direction. Spectra were approximately fully sampled by physically shifting the array by 1/2 pixel, so that two detector positions were observed per resolution element. Each spectral image was bias-subtracted and flat-fielded using a blackbody calibration source. As the background varies between the two detector positions used to sample profiles, the baselines in the resultant interlaced spectra can exhibit a sawtooth profile. If present, this is corrected for by fitting linear baselines to the odd and even pixels on either side of the emission line in each row of each spectral image respectively. Removing the odd and even pixel baselines, row by row, corrects for the ripple in the final reduced spectra. All profiles were absolutely velocity-calibrated and then corrected for the velocity of the observer relative to the dynamical Local Standard of Rest (Delhaye 1965; Gordon 1976) to an accuracy of $\leq 0.01$ km s$^{-1}$ using the Starlink RV software.

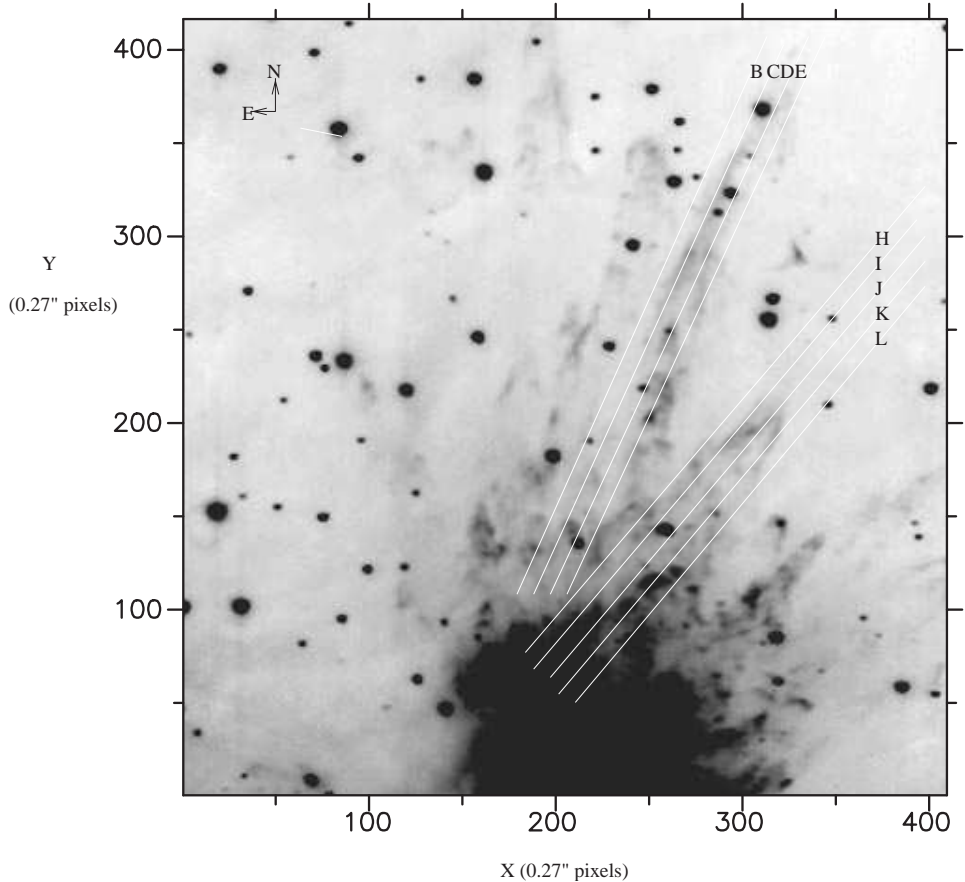

**Figure 1.** Narrow-band image in the $H_2$ 1–0 S(1) line at 2.12 μm of the northern outflow from OMC-1, consisting of 16 frames taken with the IRIS near-infrared camera on the 3.9-m Anglo-Australian Telescope (Allen & Burton 1993). Right ascension and declination axes are labelled by pixel number where the image scale is 0.27 arcsec per pixel. OMC-1 is located in the brightest region, centred on (229, 13). At least 20 hollow structures or 'wakes' are resolved, closely resembling bow shocks. Superimposed on the image are the locations of slits B, C, D, E for M42 HH126–053 and slits H, I, J, K, L for M42 HH120–114 with absolute positions listed in Table 1. Compact [Fe II] emission knots coincident with the tips of these wakes can be seen in Fig. 2.

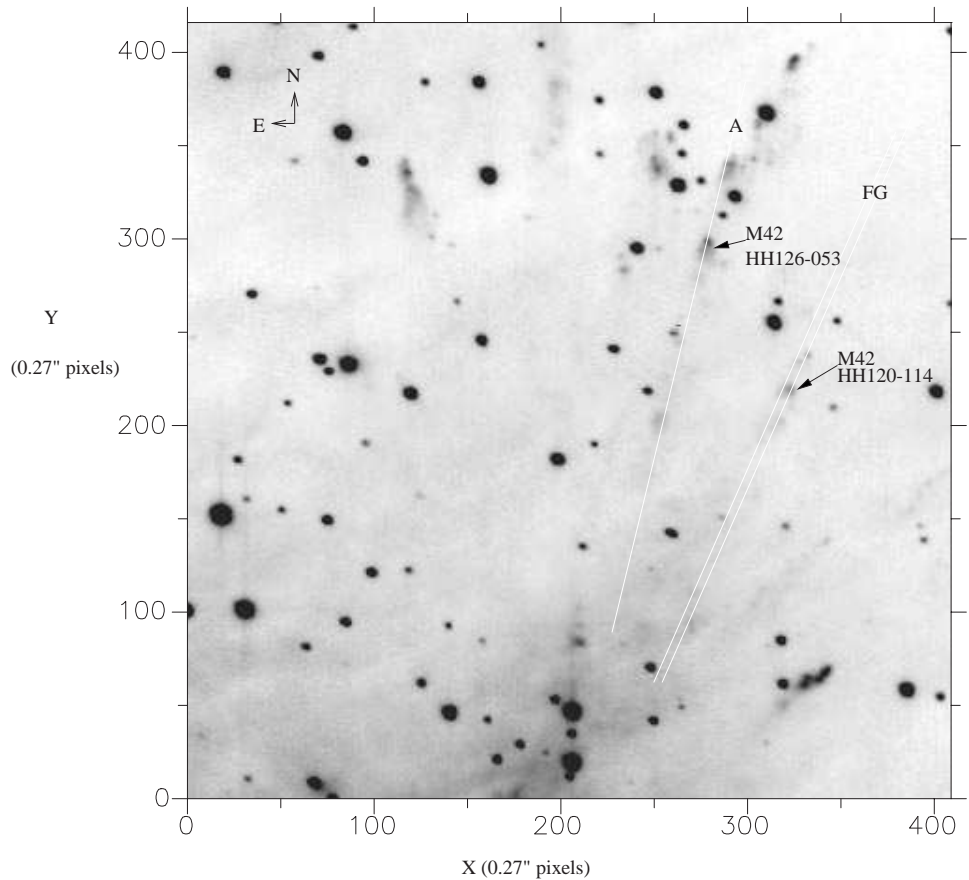

**Figure 2.** As Fig. 1, but now imaged in the [Fe II] $a^4F_{5/2}$–$a^4D_{7/2}$ line at 1.64 μm (Allen & Burton 1993). The 'bullets' appear as knots of bright emission coincident with the tips of the H$_2$ wakes in Fig. 1. Much of the diffuse nebulosity is due to the 12–4 transition of H I which also lies within this filter bandpass. Superimposed on the image is the approximate location of slit A for M42 HH126–053 and slits F, G for M42 HH120–114 with absolute positions listed in Table 1.

The [Fe II] and H$_2$ 1–0 S(1) profile observations of the bullets M42 HH126–053 and M42 HH120–114 required between five and nine groups of object–sky–sky–object observations respectively for each position. Slits were oriented along the axis of each bullet and wake, and then positioned by small step offsets from the nearby infrared reference objects, BN and IRc9, the positions of which are exactly known. In addition, pointing checks were made at approximately hourly intervals by offsetting to three nearby stars and measuring any drift in RA or Dec. from the expected position. The drift was found to be typically $\lesssim 0.5$ arcsec and rarely exceeded 1.0 arcsec. Table 1 lists the position, orientation, sky offset and pixel scale of each slit. The large offset sky positions were necessary in order to be clear of M42 nebular emission.

Wavelength calibration was achieved by measurement of arc lamp emission lines in each different instrumental configuration. For the [Fe II] observations an argon arc lamp was employed and the emission lines at $\lambda_{air} = 1.6441$ and 1.6524 μm were measured. OH emission lines present in the sky spectra at $\lambda_{air} = 1.6415$ and 1.6442 μm were also used to check the wavelength calibration. The same procedure was followed for wavelength calibration of the H$_2$ observations, using a krypton arc lamp in 26th order and measuring the emission lines at $\lambda_{air} = 2.1215$ and 2.12391 μm. For the [Fe II] observations of the bullet M42 HH126–053 made using the 58×62 pixel array, it was not necessary to correct for optical distortions as they are not resolved. However, it was necessary to rebin all image, sky and arc lamp frames made using the 256 × 256 pixel array to correct for small optical distortions in both the spatial and dispersive directions (measured on arc frames) using routines in the Starlink FIGARO software library. The effective velocity resolution for each instrumental configuration was then determined by measurement of the FWHM of the best Gaussian fit to unresolved arc lines and OH sky lines. Changes in spectral and spatial resolution after rebinning are accounted for and are negligible since the profiles are super-resolved and any broadening adds in quadrature. For the H$_2$ 1–0 S(1) profile observations, the effective velocity resolution was determined to be $23.1 \pm 0.3$ km s$^{-1}$; results for the [Fe II] observations are given in Table 2.

The nearby UKIRT standard star BS 1937 ($K = 4.47$ mag) was also observed at the same airmass as Orion to flux-calibrate the data. No stellar absorption features are present in this wavelength range. A different standard star (BS 1784) was observed for the earlier M42 HH126–053 [Fe II] observations but was later determined to be dominated by stellar absorption features at this wavelength and so was not used. A relative, rather than an absolute, flux calibration was made in this case. The angular size projected on the sky of a single CGS4 pixel in each configuration was measured by sliding a star by 20 pixels along the slit and measuring the resultant shift in declination (Table 1). An artificial, instrumental 'ghost' feature was newly identified on both arc lamp and object spectra when using the 256 × 256 pixel array. It was measured to be a blueshifted secondary at $-51.9 \pm 9.5$ km s$^{-1}$ relative to the primary line, and to have a flux of 4.5 per cent relative to the primary (real) line. It was therefore only detectable at the brighter on-wake positions and was accounted for during line fitting.

**Table 2.** Table of bullet dynamics as derived from measurements of their integrated [Fe II] 1.644-μm line emission and after subtraction of background emission. *SM* is the FWHM of the smoothing Gaussian, and differs for each bullet because of the different detector arrays used in each case. $v_{max}(0.1)$ and $v_{min}(0.1)$ are the maximum and minimum absolute velocities at 0.1 of the observed peak intensities, giving $v_{max}$ and $v_{min}$, the maximum and minimum velocities of the emitting gas after correction for the effects of thermal and instrumental broadening. $v_s$ and $\alpha$ are the resultant bullet velocity and orientation to the line of sight.

| Bullet | $SM$ /km s$^{-1}$ | $v_{max}(0.1)$ /km s$^{-1}$ | $v_{max}$ /km s$^{-1}$ | $v_{min}(0.1)$ /km s$^{-1}$ | $v_{min}$ /km s$^{-1}$ | $v_s$ /km s$^{-1}$ | $\alpha$ /° |
|---|---|---|---|---|---|---|---|
| M42 HH126–053 | 23.4±0.4 | +60±10 | +50±10 | −110±10 | −100±10 | 150±15 | 70±15 |
| M42 HH120–114 | 21.7±0.4 | +40±10 | +30±10 | −100±10 | −90±10 | 120±15 | 60±15 |

The $H_2$ trails behind the bullets studied in Orion are of the order of 40 arcsec long and 10 arcsec wide, so it was necessary to observe at five slit positions per wake in order that resultant velocity profiles could be compared for a range of positions both along and across the wakes to search for the predicted changes within the proposed bow shock structure.

## 4 RESULTS

We present observations of [Fe II] profiles in the bullets at the tips of the $H_2$ wakes, and derive their velocity and orientation, for the two bullets M42 HH126–053 and M42 HH120–114. We then present observations of the $H_2$ emission-line profiles within the associated wakes. We first introduce a standard model for a bow-shaped shock driven by a bullet, as necessary to interpret the following [Fe II] profiles.

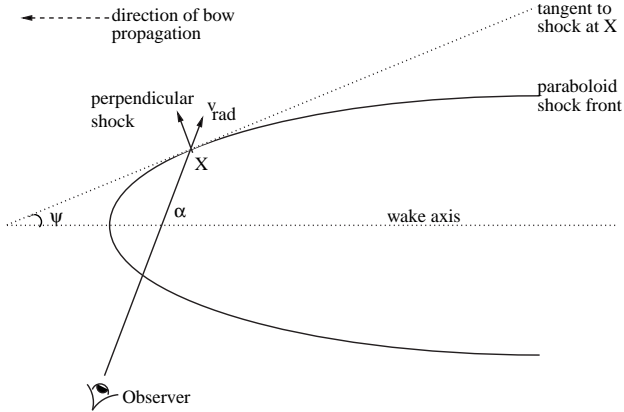

**Figure 3.** Diagram of a theoretical bow shock front (assumed paraboloid) passing at velocity $v_s$ through a stationary medium and imparting a velocity $v_\perp = v_s \sin\psi$ perpendicular to the bow at X. The observer sees an effective radial velocity $v_{rad}$ (see text). The observer resolves an oppositely directed component of shocked gas (not shown here) from the intersect of the line of sight to the near side of the wake, given the shock velocities measured for the bullet.

### 4.1 [Fe II] 1.644-μm profiles

#### 4.1.1 Derivation of bullet velocity and orientation

The two most important parameters in modelling integrated bow shock line profiles are the incident bullet velocity, $v_s$, and the angle between the axis and the line of sight, $\alpha$ (Fig. 3). Observationally, we can directly measure $v_s$ since it is equal to the FWZI of the integrated emission over an entire bow shock front, independently of the shape of the bow shock, the orientation angle, the pre-shock density, the elemental abundances and the reddening (Hartigan, Curiel & Raymond 1989). Furthermore, by measuring the maximum ($v_{max}$) and minimum ($v_{min}$) radial velocities for the integrated line profile, we obtain both $v_s$ and $\alpha$. The appropriate formulae for the observed maximum and minimum radial velocities for a bullet ploughing into a stationary medium are given by Hartigan et al. (1987) as

$$v_{max} = \frac{v_s}{2}(1 - \cos\alpha) \qquad (1)$$

and

$$v_{min} = -\frac{v_s}{2}(1 + \cos\alpha). \qquad (2)$$

The total velocity range, i.e. full width at zero intensity (FWZI = $v_{max} - v_{min}$), is evidently $v_s$. In addition, the effects of thermal and instrumental broadening are accounted for by setting the values of $v_{max}$ and $v_{min}$ to

$$v_{max} = v_{max}(0.1) - \frac{SM}{2} \qquad (3)$$

and

$$v_{min} = v_{min}(0.1) + \frac{SM}{2}, \qquad (4)$$

where $v_{max}(0.1)$ and $v_{min}(0.1)$ are the maximum and minimum velocities at 0.1 of the observed peak intensity, and *SM* is the FWHM of the smoothing Gaussian (23.4 km s$^{-1}$ at 1.644 μm for the 58 × 62 pixel array). The velocities at 0.1 of the peak intensity are measured because the detector response below this level is not perfectly Gaussian in the profile tails and would distort the resultant Gaussian width calculated after correction for smoothing.

#### 4.1.2 M42 HH126–053 [Fe II] 1.644-μm profiles

Observations of the [Fe II] 1.644-μm line profiles in the bullet M42 HH126–053 (slit A, Table 1), shown in Fig. 4(a) alongside its corresponding $H_2$ wake (Fig. 4b), apparently oriented in or close to the plane of sky, are displayed in Fig. 5. The profiles are broad and strongly peaked at a velocity of $\simeq 8$ km s$^{-1}$ measured relative to the dynamical Local Standard of Rest, close to the rest velocity of the ambient medium of $+9\,\mathrm{km\,s^{-1}}$ determined by Goldsmith, Plambeck & Chiao (1975) from $^{10}$CO and $^{13}$CO radio observations. It can be seen that the total velocity range is highest at the leading edge of the emission (top rows), as is expected at the head of a bow shock where the gas is shocked at velocities approaching the full driving velocity of the bullet in two components, oppositely directed from one another. Further down the bullet, the velocity range narrows (bottom of Fig. 5) as the ambient gas sees a lower effective shock velocity (the normal component of the shock velocity decreases moving down each side of the bow-shaped shock front geometry).

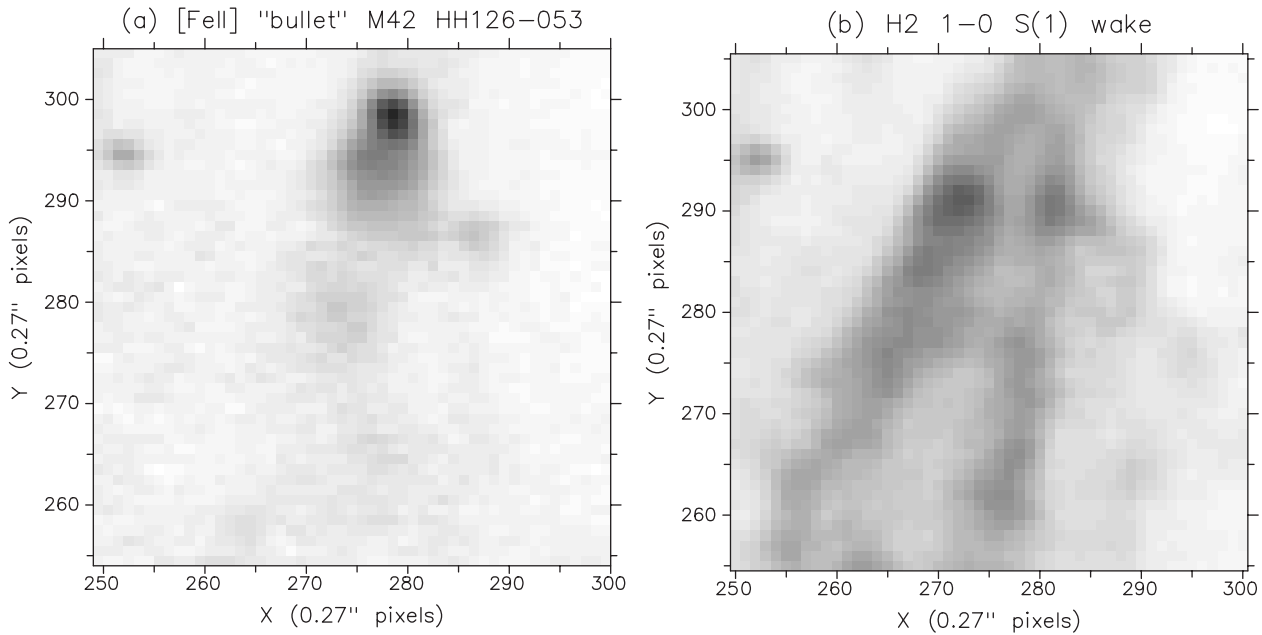

**Figure 4.** A close-up of the M42 HH126–053 bullet region in (a) [Fe II] 1.644-μm emission, and (b) $H_2$ 1–0 S(1) emission from the images of Allen & Burton (1993) with a 0.27-arcsec pixel scale.

In this case, applying equations (3) and (4), the integrated [Fe II] profile (Fig. 6) corresponds to a shock velocity of $v_s = 150 \pm 15\,\mathrm{km\,s^{-1}}$ and an orientation angle of $\alpha = 70° \pm 15°$ to the line of sight (Table 2). We assume that we observe the maximum velocity range in the bullet emission, given that the slit position fully samples the brightest emission at the tip of the bullet where shock velocities are highest. The weaker, downstream emission not sampled by our slit can therefore only alter the shape of the integrated profile and not the measured values of $v_{\max}(0.1)$ and $v_{\min}(0.1)$. Note that $\alpha = 90°$ for a bow shock oriented exactly in the plane of the sky. A comparison of this observed [Fe II] profile shape with radiative bow shock model predictions (Hartigan et al. 1989) is hampered in this case because of the difficulty of subtracting a background emission component (not clearly identified for this observation made using the older, less sensitive 58×62 pixel array). An unambiguous identification of this component in the following section, however, indicates that the observed values of $v_{\min}$ and $v_{\max}$ above will remain unaltered, and the resultant shock velocity and orientation therefore stand. Encouragingly, the observed profile in Fig. 5 is similar in shape to the theoretical integrated velocity profile generated for a $v_s = 200\,\mathrm{km\,s^{-1}}$, $\alpha = 60°$ bullet (Hartigan et al. 1987), with a suitably reduced FWZI velocity range.

### 4.1.3 M42 HH120–114 [Fe II] 1.644-μm profiles

The bullet M42 HH120–114 is shown in Fig. 7(a), alongside its corresponding $H_2$ wake in Fig. 7(b). Observations of the [Fe II] 1.644-μm line profiles at two adjacent slit positions (F and G) including the bullet M42 HH120–114 are shown in Fig. 8, superimposed on the Allen & Burton (1993) [Fe II] image. Each profile is displayed in a box centred on the exact position of the corresponding CGS4 slit row. The common flux and velocity scales of each profile are illustrated in the 'key' box. This method of displaying the relative flux and velocity scales of profiles for a given position on a bullet/wake in the region will be used throughout the remainder of this paper. As for M42 HH126–053, bullet profiles are broad and strongly peaked at a velocity of $\simeq +8\,\mathrm{km\,s^{-1}}$, close to the rest velocity of the ambient medium of $+9\,\mathrm{km\,s^{-1}}$. Again, the emission profiles narrow as one moves downstream from each individual bullet tip, such that the broadest profiles are associated with the brightest emission at the tips of the bullets, and the wake is apparently oriented in or close to the plane of the sky.

In addition to the broadened velocity profiles associated with the bullet, it is immediately apparent that the entire region covered by our slit positions is pervaded by a constant 'background' emission feature. At rows well to the north and south of the bullet positions, the profile is dominated by the narrow background component centred at zero velocity. We also note that it is possible to discern other bullet-like features in Fig. 7(a) close to the prominent M42 HH120–114 bullet at (323, 218). Indeed, a weak [Fe II] knot centred at (318, 203) may be associated with a second $H_2$ wake that partially overlaps with that of M42 HH120–114 (Fig. 7b).

In order to determine accurately the velocity range of emission resulting from the bullet alone, it is necessary to isolate the background emission pervading the entire region. For this purpose, slits F and G (Fig. 8) were positioned so that a significant area upstream of the bullet emission was sampled. The averaged upstream profile was found by co-adding all observed profiles upstream of the bullets at these positions, and is displayed in Fig. 9. The background profile is centred at $-0.4 \pm 0.2\,\mathrm{km\,s^{-1}}$ (weighted mean) and is well fitted by a single Gaussian profile of FWHM = $31.1 \pm 0.6\,\mathrm{km\,s^{-1}}$. After deconvolution of the instrumental profile (FWHM = $21.7 \pm 0.4\,\mathrm{km\,s^{-1}}$ for the $256 \times 256$ pixel array at 1.644 μm), the intrinsic FWHM is $22.3 \pm 0.7\,\mathrm{km\,s^{-1}}$. The line flux is $(3.4 \pm 0.2) \times 10^{-18}\,\mathrm{W\,m^{-2}}$ in a $1.69 \times 0.9\,\mathrm{arcsec^2}$ CGS4 pixel row.

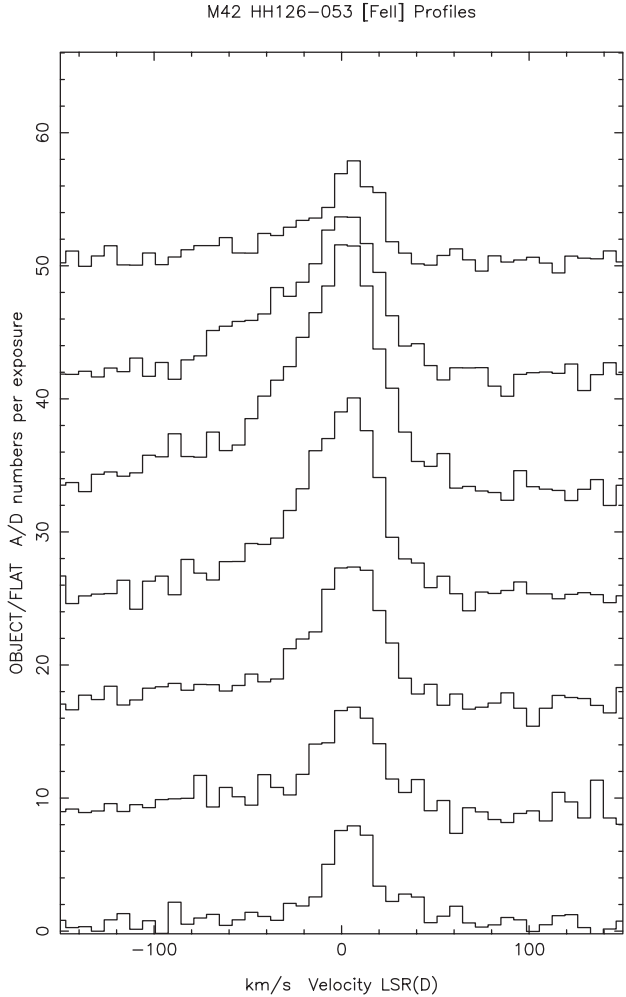

**Figure 5.** M42 HH126–053 [Fe II] 1.644-μm velocity profiles (slit A in Table 1), running from row 19 (top) to row 25 (bottom) at 2.16-arcsec spacing along the bullet axis with peak intensity at the bullet tip. Profiles are relatively but not absolutely flux-calibrated in this case (see text).

This averaged background profile was therefore subtracted from each observed profile in order to isolate bullet-only emission. Fig. 10 shows the bullet-only emission profiles in slits F and G coincident with the bullet. The resultant profiles consist of at least two components, as expected for bow shock emission profiles. Variations in background emission show as relatively small spikes and dips at zero velocity if significantly different from the average. The integrated profiles, summed over all positions including the M42 HH120–114 bullet, are shown in Fig. 11 (total observed emission) and Fig. 12 (background component subtracted). Even after subtraction of the background, it is clear that the strongest peak emission from the bullet lies within $\sim 40\,\mathrm{km\,s^{-1}}$ of zero radial velocity. This is consistent with bullet models, as opposed to models in which the molecular gas flows past an obstacle and the peak emission lies at relatively higher velocities for a given incident shock velocity (Hartigan et al. 1987).

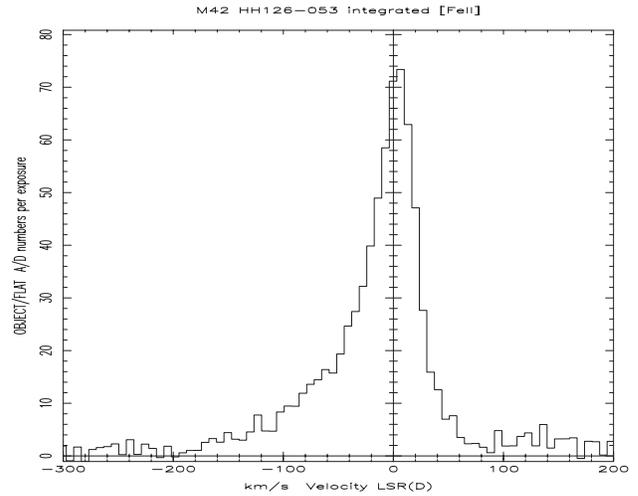

**Figure 6.** M42 HH126–053 integrated [Fe II] 1.644-μm velocity profile for positions in slit A (Table 1) coincident with the bullet. The profile shows the maximum and minimum absolute velocities and thus the total velocity range in the bullet.

Using the methods described previously, the background-subtracted integrated [Fe II] profile for the bullet M42 HH120–114 (Fig. 12) corresponds to a shock velocity of $v_s = 120 \pm 10\,\mathrm{km\,s^{-1}}$ and an orientation angle of $\alpha = 60° \pm 15°$ to the line of sight, after small corrections for thermal and instrumental broadening (Table 2). A comparison of this observed [Fe II] profile shape with radiative bow shock model predictions (Hartigan et al. 1987; 1989) shows good agreement with the theoretical integrated velocity profile generated for a $v_s = 100\,\mathrm{km\,s^{-1}}$, $\alpha = 60°$ bow shock. In particular, peak emission is blueshifted by up to $\sim 40\,\mathrm{km\,s^{-1}}$ from the rest velocity for the observed profile, consistent with a similar but slightly larger blueshift of $\sim 50\,\mathrm{km\,s^{-1}}$ from rest for the theoretical profile.

### 4.2 Shocked H₂ 1–0 S(1) profiles in associated bullet wakes

As with the [Fe II] emission in M42 HH120–114, a roughly constant emission feature is seen at all positions upstream and adjacent to the wakes, while profile intensity increases by a factor of 10 or more up to a maximum of $1.68 \pm 0.01 \times 10^{-16}\,\mathrm{W\,m^{-2}}$ in a $1.74 \times 0.9\,\mathrm{arcsec^2}$ CGS4 pixel at positions on the wakes themselves. The emission profiles are dominated at all positions by a broad, single peak centred at or close to zero velocity relative to the dynamical Local Standard of Rest. Before analysing the profiles in more detail, it is necessary to determine and subtract the constant background feature, as for the [Fe II] profiles in M42 HH120–114.

#### 4.2.1 H₂ background component

The averaged, upstream H₂ 1–0 S(1) profile was found by averaging profiles at positions well clear of the wake features. The consequent profiles are almost identical for both regions. Background flux variations are of the order of $\lesssim 10^{-18}$ W m$^{-2}$, which is very small compared with the on-wake fluxes ($\sim 10^{-16}$ W m$^{-2}$). The majority of profiles used in this determination are at positions far upstream of the M42 HH120–114 wake. The resultant profile is shown in Fig. 13, and is well fitted by a single Gaussian centred at $2.5 \pm 0.5$ km s$^{-1}$, close to the value of $+4$ km s$^{-1}$ found by Moorhouse et al. (1990) for H$_2$ 1–0 S(1) profiles in the OMC-1 region, and in close agreement with that determined for the [Fe II] background component. The FWHM is $34.0 \pm 2.5$ km s$^{-1}$. After deconvolution of instrumental broadening (FWHM $= 23.1 \pm 0.3$ km s$^{-1}$), the intrinsic FWHM of the background component is therefore $26.0 \pm 2.5$ km s$^{-1}$, only slightly higher than for the [Fe II] 1.644-µm background component, i.e. within $1\sigma$ for each. The average background flux in the H$_2$ 1–0 S(1) transition is $9.9 \pm 0.6 \times 10^{-18}$ W m$^{-2}$ in a $1.74 \times 0.9$ arcsec$^2$ CGS4 pixel.

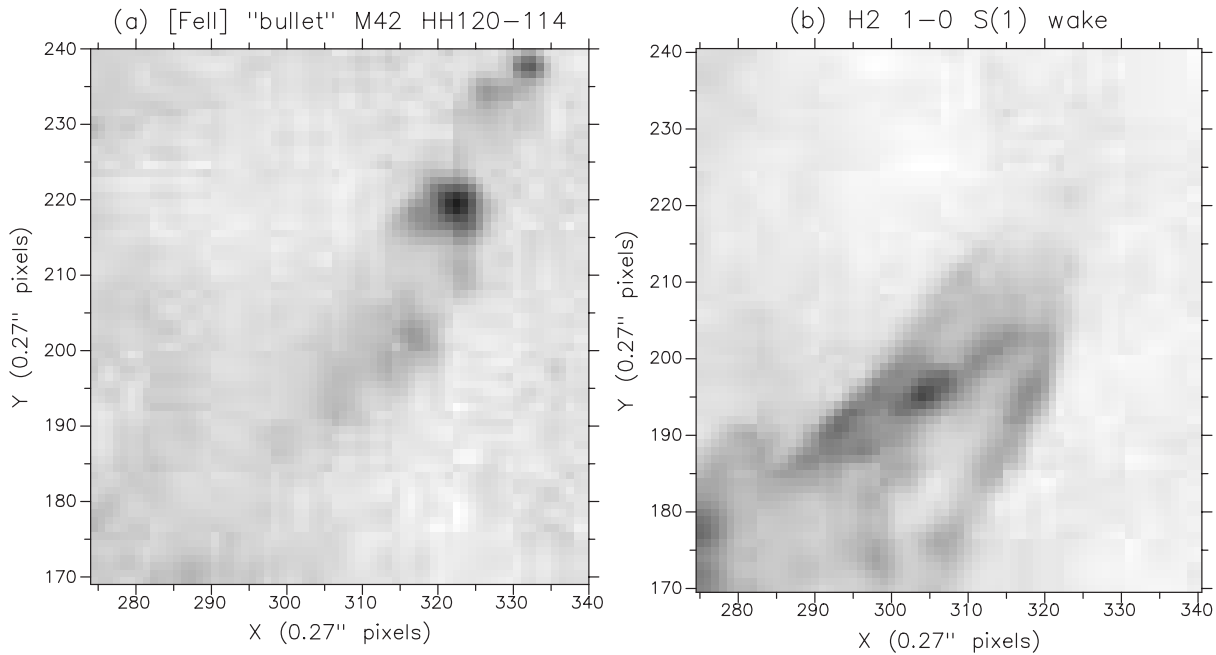

**Figure 7.** A close-up of the M42 HH120–114 bullet region in (a) [Fe II] 1.644-µm emission and (b) H$_2$ 1–0 S(1) emission from the images of Allen & Burton (1993) with a 0.27-arcsec pixel scale.

### 4.2.2 M42 HH126–053 H$_2$ 1–0 S(1) wake profiles

The averaged background component (Fig. 13) was subtracted from all positions on the H$_2$ wakes to determine the characteristics of the wake-only emission. Fig. 14 shows the resultant profiles superimposed on the H$_2$ wake of M42 HH126–053 (slits B, C, D, E). Profile intensity increases for on-wake positions up to a maximum of $1.58 \pm 0.01 \times 10^{-16}$ W m$^{-2}$ in a $1.74 \times 0.9$ arcsec$^2$ CGS4 pixel (after background subtraction). Detailed Gaussian line fits of peak velocity and FWHM were possible for the emission profiles at all positions. Routines in the SPECDRE Starlink software library were employed and the results, including intrinsic FWHM after deconvolution of instrumental broadening, are tabulated in Table 3 for the main peak emission centred at or close to zero velocity. For each slit position, a small but significant blueshift is observed up to a peak velocity of $-4.9 \pm 1.0$ km s$^{-1}$ with respect to the dynamical Local Standard of Rest, compared with the off-wake peak of $+2.5 \pm 0.5$ km s$^{-1}$, corresponding to a blueshift of magnitude $7.4 \pm 1.1$ km s$^{-1}$ and up to $14$ km s$^{-1}$ from the ambient cloud velocity of $+9$ km s$^{-1}$ in this region measured in CO (Goldsmith et al. 1975). The maximum blueshift occurs at positions corresponding to the brightest regions at the tips of the H$_2$ emission, and almost coincident (along the line of sight) with the [Fe II] 1.644-µm bullet. In moving down the wake, profiles weaken and the velocity of the peak approaches the velocity of the background once again. At no point on the wake is there any evidence of the double-peaked profile structure in the dominant emission profile centred near zero velocity, such as would be expected for bow shocks.

The decrease in both the intensity and blueshift of the profiles in moving down the central on-axis wake positions is coincident with a corresponding decrease in the intrinsic (i.e. with instrumental broadening deconvolved) FWHM of the profiles, ranging from a maximum of $24.3 \pm 0.4$ km s$^{-1}$ near the tip of the wake down to $10.0 \pm 0.6$ km s$^{-1}$ at the tail of the wake for the two on-wake slits C and D (Table 3). Slit B lies along the edge of the main wake emission associated with M42 HH126–053, and shows a less clear trend in profile that is perhaps due to the irregular shape of the underlying wake geometry. Slit E follows the general trend of the central positions but appears to be coincident with an additional wake feature corresponding to a secondary [Fe II] clump (Fig. 4) which confuses interpretation.

Additional and much weaker components are identified at velocities significantly blueshifted from the dominant single peak and highlighted in Fig. 15. These features are strongest at the correspondingly brightest main peak positions. One of these components remains fairly constant and is identified as the 'ghost' feature previously observed in arc lamp spectra and centred at about $-51$ km s$^{-1}$, and the intensity of which is proportional to the main peak intensity and up to $\sim 5$ per cent in relative flux intensity, consistent with the arc line fits. However, a component centred at velocities as great as $-106 \pm 3$ km s$^{-1}$ is also observed and is often stronger than the ghost feature. Results of Gaussian line fitting to any weak component that is clearly distinguished from the constant ghost feature associated with the much brighter, zero-velocity centred component are shown in Table 4. It is noted that both the peak velocities and the intrinsic FWHM of these components follow the trends found for the main peak emission as one moves down and outwards from the tip of the wake, although accurate determinations are difficult because of the much lower signal-to-noise ratio.

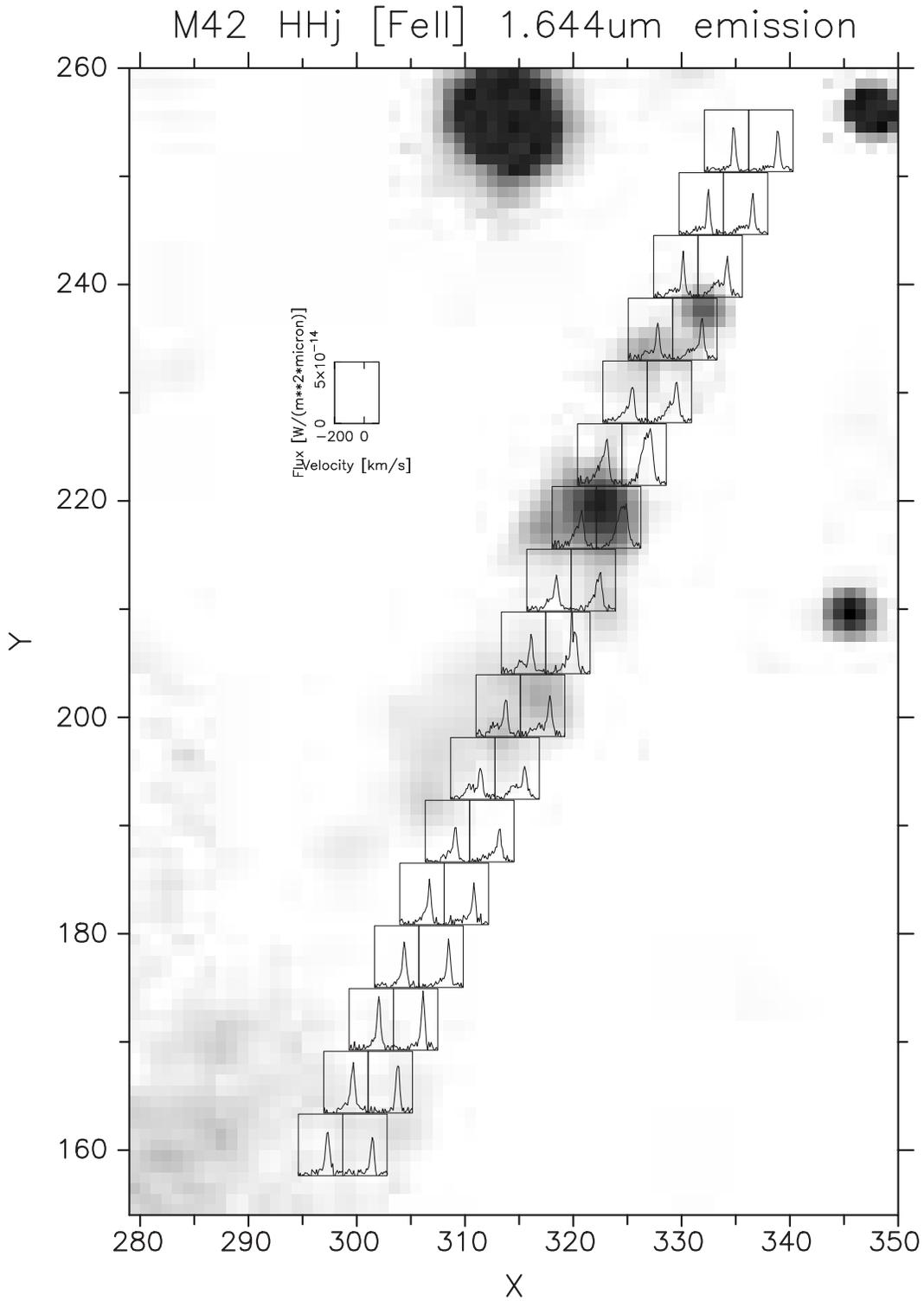

**Figure 8.** The region including the M42 HH120–114 bullet centred at (323, 218) imaged in [Fe II] 1.644 µm (Allen & Burton 1993, 0.27-arcsec pixel scale), with the corresponding [Fe II] velocity profiles at two slit positions superimposed. Referring to Table 1, slit row 25 (for which RA and Dec. are recorded) is at image coordinates (323, 225) and (327, 225) for slits F and G respectively. Each profile box is centred on the appropriate CGS4 pixel row and has equal flux and velocity scale as indicated by the key. Note the constant and relatively narrow background component in addition to the bullet emission.

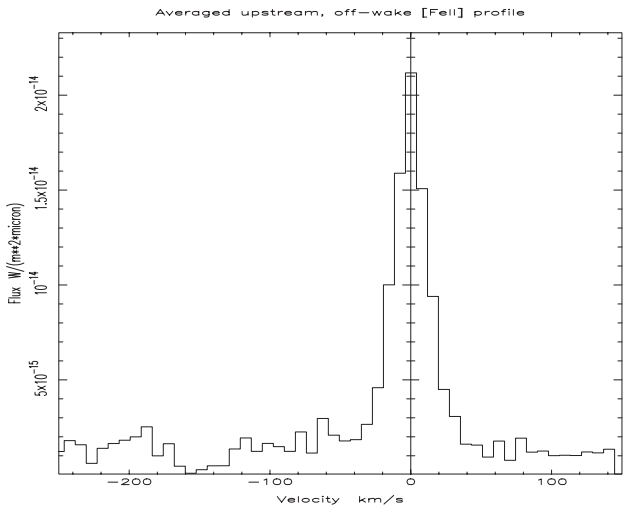

**Figure 9.** Averaged background [Fe II] 1.644-μm velocity profile, upstream of M42 HH120–114.

### 4.2.3 M42 HH120–114 $H_2$ 1–0 S(1) wake profiles

Fig. 16 shows the background-subtracted $H_2$ 1–0 S(1) profiles superimposed on the $H_2$ wake of M42 HH120–114 (slits H, I, J, K, L). As for M42 HH126–053, emission is dominated by a broad but singly peaked profile. The profile intensity increases for on-wake positions up to a maximum of $1.21 \pm 0.02 \times 10^{-16}$ W m$^{-2}$ in a $1.74 \times 0.9$ arcsec$^2$ CGS4 pixel after subtraction of the averaged background intensity. Detailed Gaussian line fits of peak velocity and FWHM were made to the emission profiles as before, and the resultant peak velocities and deconvolved FWHM values are tabulated in Table 5 for the main peak emission centred at or close to zero velocity. Once again, at no point on the wake is there any evidence of the double-peaked profile structure in the dominant emission profile centred near zero velocity, as would be expected for bow shocks.

Small peak velocity shifts are again observed. However, in this case, the peaks are slightly redshifted compared with the background emission peak. For each slit position, a small but significant redshift is observed for the peak of up to $10.5 \pm 0.1$ km s$^{-1}$ compared with the average off-wake peak of $2.5 \pm 0.5$ km s$^{-1}$, corresponding to a redshift of up to $8.0 \pm 0.6$ km s$^{-1}$. The largest shifts from row to row again occur at the head of the wake and are again almost coincident (along the line of sight) with the main [Fe II] 1.644-μm bullet emission. Although less clearly defined than for M42 HH126–053, since for positions westward of slit I a second bullet and wake significantly overlap, there is a clear trend for the peak velocities to reduce from the maximum observed deflection and approach the background peak velocity once again in moving down the wake.

The decrease in intensity and redshift of the profiles at the central on-axis wake positions of the main wake M42 HH120–114 is again coincident with a corresponding decrease in the intrinsic FWHM of the profiles, in this case ranging from a maximum of typically $26.1 \pm 2.4$ km s$^{-1}$ near the leading edge of the wake down to $13.5 \pm 0.5$ km s$^{-1}$ at the tail of the wake for slits I, J, K. We note, however, a significantly higher FWHM at the lowest row numbers 19 and 20 (at the tip of the wake) than is seen elsewhere. This is due to a very weak but broad pedestal in addition to the weak main profile, which biases the Gaussian line fits to give a resultant FWHM as high as $91.5 \pm 10.9$ km s$^{-1}$ for row 19 of slit L. Also, at slit K, for example, the FWHM of the line emission appears to rise and fall twice with increasing row number, correlated with the two different wakes observed along the line of sight at these positions.

Unlike the wake associated with M42 HH126–053, additional velocity components are not clearly identified on the main wake region itself, apart from the weak, artificial ghost feature. However, real secondary components are visible on the confused region just downstream of the actual wake(s). The broad but very weak emission noted above for the lowest row numbers in Fig. 16 is, however, visible. At these positions, ill-defined in the $H_2$ image but coincident with the strongest [Fe II] bullet emission, it appears that the total velocity range at close to zero intensity is similar to that of the corresponding individual [Fe II] 1.644-μm profile, i.e. as high as $130 \pm 20$ km s$^{-1}$ at FWZI. Fig. 17 shows row 19 of slit K but with the background not subtracted to show the weak high-velocity emission at higher signal-to-noise ratio. We note the similarity in both shape and velocity range to the corresponding [Fe II] bullet profile in Fig. 8 at the same position. Retention of the background component does not contribute significantly to the high-velocity emission in which we are interested here. The small dip near the peak of the profile is artificially introduced by a very weak and unresolved $H_2$ emission source centred at zero velocity in this particular sky position used, and so again will not affect the high-velocity emission. This sky feature was carefully checked for all relevant positions in this wake, and confirmed to be approximately constant but negligible in intensity compared with even the background component.

## 5 DISCUSSION

We first describe the historical development of our understanding of the thermally excited $H_2$ emission in Orion, as a necessary prerequisite to the following confrontation with our new data set. We consider the various steady state-shock models, shock front geometries and the possible effects of instabilities and dust. In each case, the future work required to tackle the outstanding issues raised is defined.

### 5.1 History of $H_2$ observations in Orion

In 1979, Nadeau & Geballe published high spectral resolution ($\sim 20$ km s$^{-1}$) velocity profiles of the $H_2$ 1–0 S(1) profiles in this region. After deconvolution of the instrumental profile, the intrinsic full width at half-maximum (FWHM) velocities of the lines was found to lie in the range 18 to 58 km s$^{-1}$. Nadeau & Geballe deduced two separate components to the profiles (depending on position) at this resolution having expansion velocities from BN–KL of $\sim 40\,\mathrm{km\,s^{-1}}$, and a much smaller component at $\sim 100\,\mathrm{km\,s^{-1}}$. Hence it was clear that the excitation of the observed $H_2$ could not be caused by a single plane shock front moving at or below the dissociation speed limit of $H_2$ for hydrodynamic J-shocks of $v_s \lesssim 24\,\mathrm{km\,s^{-1}}$ (Kwan 1977). Furthermore, the observed asymmetries of the line profiles were shown by Beckwith, Persson & Neugebauer (1979) to be inconsistent with models having variable extinction across the source alone. The magnitude of foreground extinction, however, was revised significantly downwards by Scoville et al. (1982) to $A_{2.1\mu m} \simeq 1.2\text{--}2\,\mathrm{mag}$, and shown to vary by a factor of 2 on scales of $\sim 4\,\mathrm{arcsec}$. The average excitation temperature was confirmed as $T_\mathrm{vib} = 2010 \pm 50\,\mathrm{K}$ for the transitions measured.

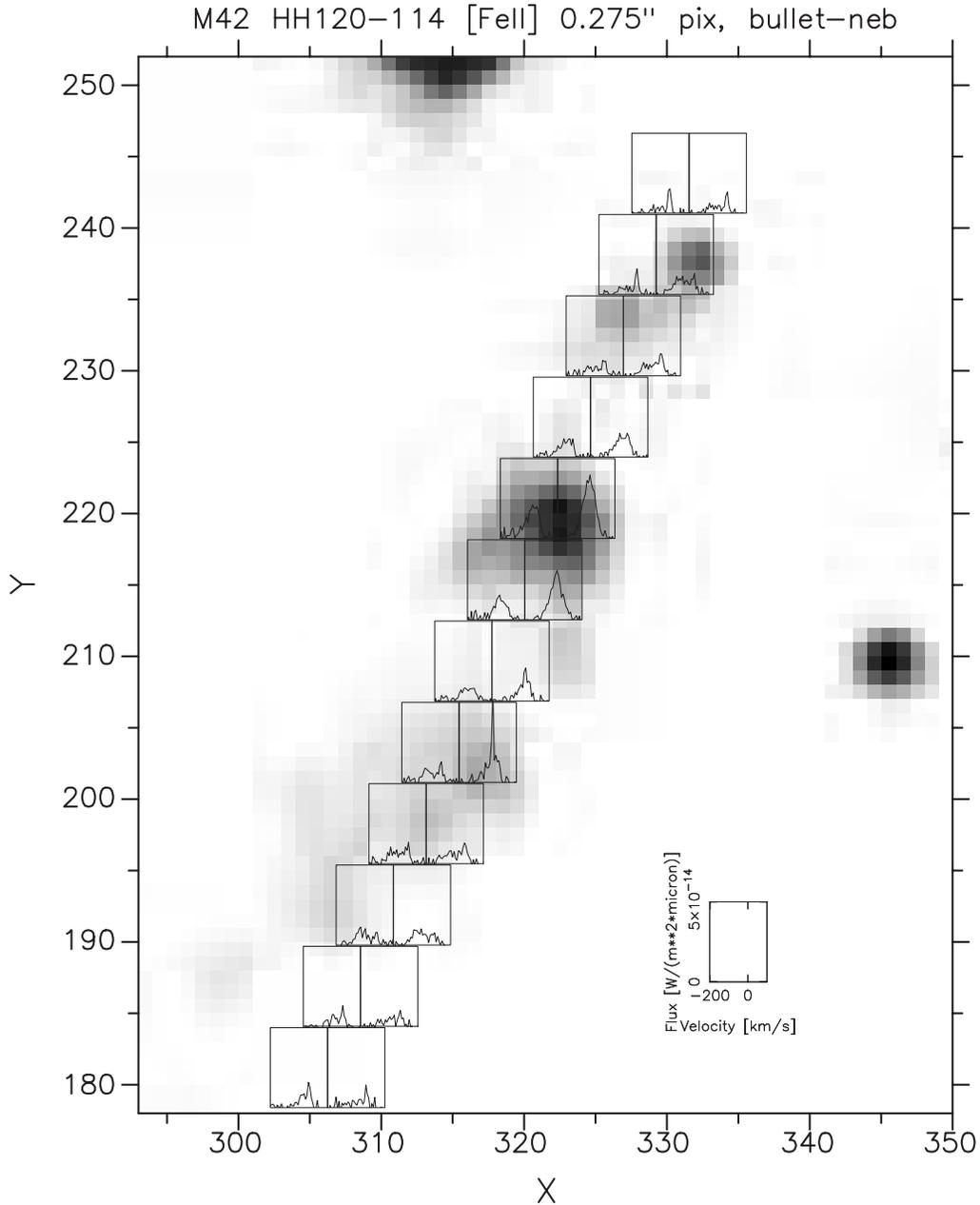

**Figure 10.** M42 HH120–114 [Fe II] 1.644-μm background-subtracted velocity profiles showing bullet-only emission.

Nadeau, Geballe & Neugebauer (1982) presented higher spatial resolution (5 arcsec) velocity profiles of the $H_2$ 1–0 S(1), 1–0 S(0) and 2–1 S(1) profiles for a range of positions about OMC-1. In addition to the broad (FWZI $\geq 100\,\mathrm{km\,s^{-1}}$), asymmetric profiles observed previously at OMC-1 itself, they noted narrower, symmetric profiles towards the periphery of the region, having an intrinsic FWHM $= 22 \pm 2\,\mathrm{km\,s^{-1}}$. The profiles were all identical in shape and found not to vary temporally. Since thermal broadening alone at $\sim 2000\,\mathrm{K}$ results in a line profile of FWHM $\leq 7\,\mathrm{km\,s^{-1}}$, it was concluded that the high-velocity $H_2$ comes from shocked gas in the flow while the low-velocity $H_2$ comes from shocked gas in the molecular cloud. Geballe et al. (1986) later ruled out scattering owing to intervening dust grains as an explanation of the broad lines, since they found the H$_2$ 1–0 O(7) line at 3.81 μm to be broader than the 1–0 S(1) line at 2.12 μm in the centre of the cloud. Scattering of photons by dust particles must decrease at longer wavelengths. First spectro-polarimetric measurements by Burton et al. (1988) found no evidence for a change in polarization across the H$_2$ 1–0 S(1) profile as would be associated with dust scattering. They proposed that the extended blue wings were produced by fast-moving clumps embedded within the outflow, while the symmetric line core resulted from shocked gas at the edges of an expanding outflow cavity.

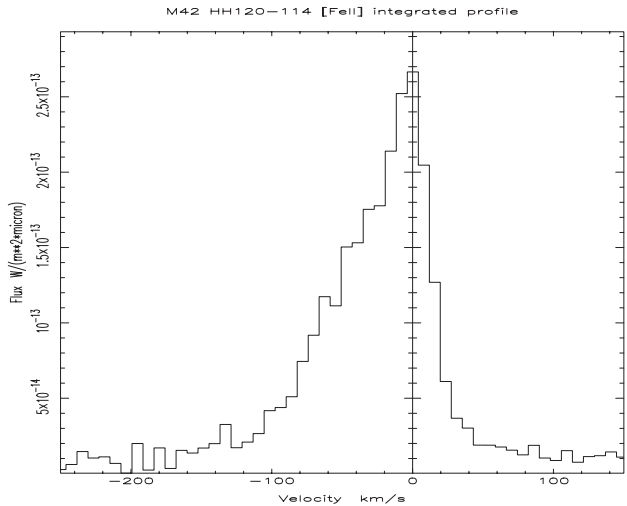

**Figure 11.** M42 HH120–114 [Fe II] 1.644-μm integrated velocity profile.

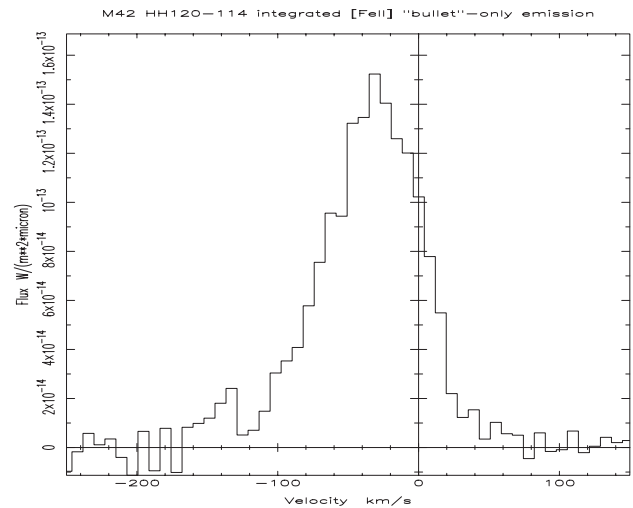

**Figure 12.** M42 HH120–114 [Fe II] 1.644-μm integrated bullet-only velocity profile (background subtracted).

Later 12 km s$^{-1}$ channel resolution observations of the H$_2$ 1–0 S(1) line profile in OMC-1 were made through a 5-arcsec beam by Brand et al. (1989b), and it was found to be smooth with an enhanced blue wing and FWZI of 140 km s$^{-1}$. There was no evidence of substructure within the profile. As already discussed, this cannot be reconciled with any planar shocks. Even an ensemble of emitting cloudlets along the line of sight, each contributing its own velocity profile, requires an unreasonable redistribution of momentum within the source. The individual cloudlets would have to achieve velocities of $\pm 70$ km s$^{-1}$ within a cylinder 0.01 pc in diameter along the entire line of sight everywhere within the source (Brand et al. 1989b). Recently, Fabry–Perot (FP) observations at 12 km s$^{-1}$ channel resolution by Chrysostomou et al. (1997) did finally resolve doubly peaked shocked H$_2$ line profiles at certain positions which resemble those expected for bow shocks buried inside the confused morphology of the central OMC-1 region. NICMOS imaging by Stolovy et al. (1998) and spectroscopic imaging by McCaughrean & MacLow (1997) have now confirmed that many overlapping H$_2$ bows are present close to OMC-1. However, the observed line splitting cannot be unambiguously identified with single bow shocks in the images, which hampers detailed model comparisons. We now compare our new results within the most clearly resolved bow-shaped shock fronts associated with the individual [Fe II] bullets seen by Allen & Burton (1993) with the models previously described.

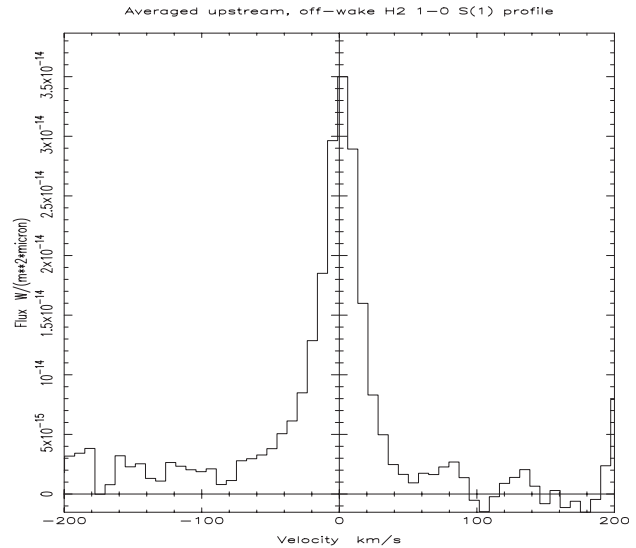

**Figure 13.** The averaged background H$_2$ 1–0 S(1) velocity profile calculated from positions significantly upstream of M42 HH120–114.

### 5.2 Comparison of H$_2$ wake profiles with steady-state bow shock models

The main aim of this investigation was to test the predictions of molecular bow shock models in the most clearly defined examples newly resolved in Orion. We have demonstrated that integrated [Fe II] 1.644-μm line profiles in Orion are entirely consistent with theoretical bow shock predictions for two different values of $v_s$ and $\alpha$ (Table 2). Motivated by clearly resolved bow-shaped H$_2$ wakes associated with the [Fe II] bullets in Orion, it was expected that observations of individual H$_2$ line profiles within the wakes should enable a clear distinction between competing models of the shock excitation within these structures.

Inconsistencies are immediately apparent between the H$_2$ and [Fe II] peak velocities at the same positions. Comparison of Figs 5 and 14 for M42 HH126–053 and Figs 10 and 16 for M42 HH120–114 shows that [Fe II] peaks and H$_2$ peaks are clearly not consistent after background subtraction in each case. The [Fe II] emission has a much larger velocity range than the $H_2$ emission, as expected given the faster shock excitation conditions of this species and the dissociation speed limit for $H_2$. This is why [Fe II] emission is seen at the heads of the bullet wakes where shock speeds are highest. [Fe II] peak velocities in M42 HH120–114, for example, are significantly blueshifted by up to $50\,\mathrm{km\,s^{-1}}$ from the ambient velocity of the cloud medium of $+9\,\mathrm{km\,s^{-1}}$ (Goldsmith et al. 1975), as expected in shock models for individual positions within a bow-shaped geometry.

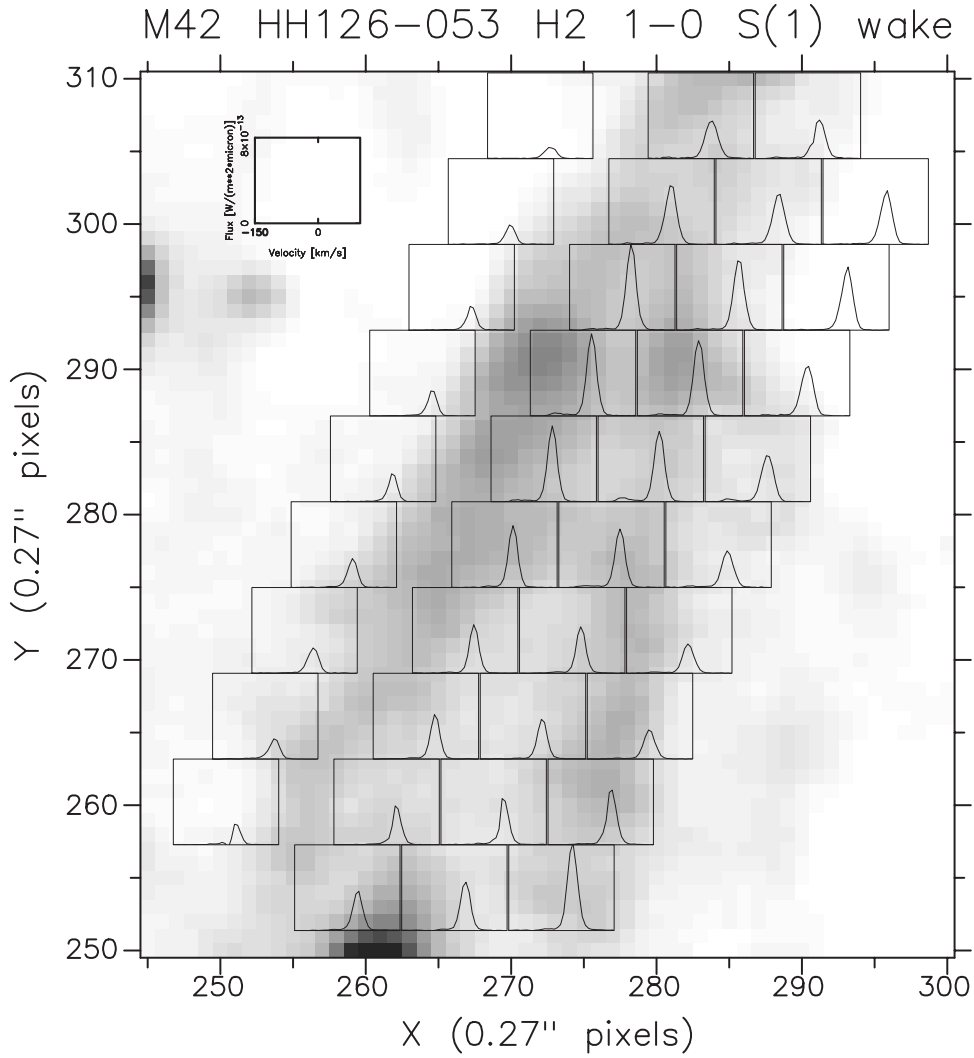

**Figure 14.** M42 HH126–053 background-subtracted $H_2$ 1–0 S(1) velocity profiles directly associated with the bullet.

Examining the morphology of the $H_2$ wakes in the Allen &

**Table 3.** Gaussian line fits to the main velocity component of the background-subtracted $H_2$ 1–0 S(1) wake profiles in M42 HH126–053. Velocities are calibrated relative to the dynamical Local Standard of Rest. 'U' indicates that the profile is unresolved.

| Slit row | Slit B $v_{peak}$ /km s⁻¹ | Slit B intrinsic FWHM /km s⁻¹ | Slit C $v_{peak}$ /km s⁻¹ | Slit C intrinsic FWHM /km s⁻¹ | Slit D $v_{peak}$ /km s⁻¹ | Slit D intrinsic FWHM /km s⁻¹ | Slit E $v_{peak}$ /km s⁻¹ | Slit E intrinsic FWHM /km s⁻¹ |
|---|---|---|---|---|---|---|---|---|
| 19 | −1.1±0.3 | 23.1±0.7 | 0.2±0.1 | 26.8±0.4 | 2.7±0.3 | 21.9±0.8 | 1.4±0.1 | 23.0±0.4 |
| 20 | −2.8±0.2 | 14.3±0.5 | −2.7±0.1 | 21.1±0.4 | −1.1±0.1 | 24.3±0.4 | 0.3±0.1 | 22.2±0.4 |
| 21 | −1.8±0.2 | 11.2±0.6 | −4.0±0.1 | 14.0±0.4 | −2.6±0.1 | 18.2±0.4 | 0.5±0.2 | 20.8±0.5 |
| 22 | −0.8±0.2 | 5.0±0.5 | −4.5±0.7 | 14.0±0.3 | −4.3±0.9 | 16.8±0.4 | −1.0±0.1 | 23.7±0.4 |
| 23 | −1.7±0.2 | U | −4.6±0.6 | 13.2±0.3 | −4.9±0.8 | 16.3±0.4 | −2.0±1.0 | 24.3±0.4 |
| 24 | −4.4±0.2 | 14.0±0.5 | −4.7±0.6 | 11.4±0.3 | −4.9±1.0 | 14.6±0.4 | −3.6±0.9 | 22.1±0.4 |
| 25 | −4.7±0.2 | 19.8±0.5 | −4.0±0.6 | 11.8±0.3 | −4.4±0.1 | 12.2±0.4 | −3.9±0.1 | 20.6±0.4 |
| 26 | −2.0±0.2 | 10.0±0.6 | −3.1±0.8 | 11.2±0.4 | −2.9±1.0 | 13.4±0.4 | −2.8±0.1 | 20.4±0.4 |
| 27 | – | – | −0.6±0.2 | U | 0.5±0.3 | U | 0.9±0.2 | 13.4±0.6 |
| 28 | −0.5±0.2 | 15.6±0.5 | −0.1±0.1 | 12.4±0.4 | 1.2±0.9 | 14.5±0.4 | 1.1±0.1 | 15.8±0.4 |

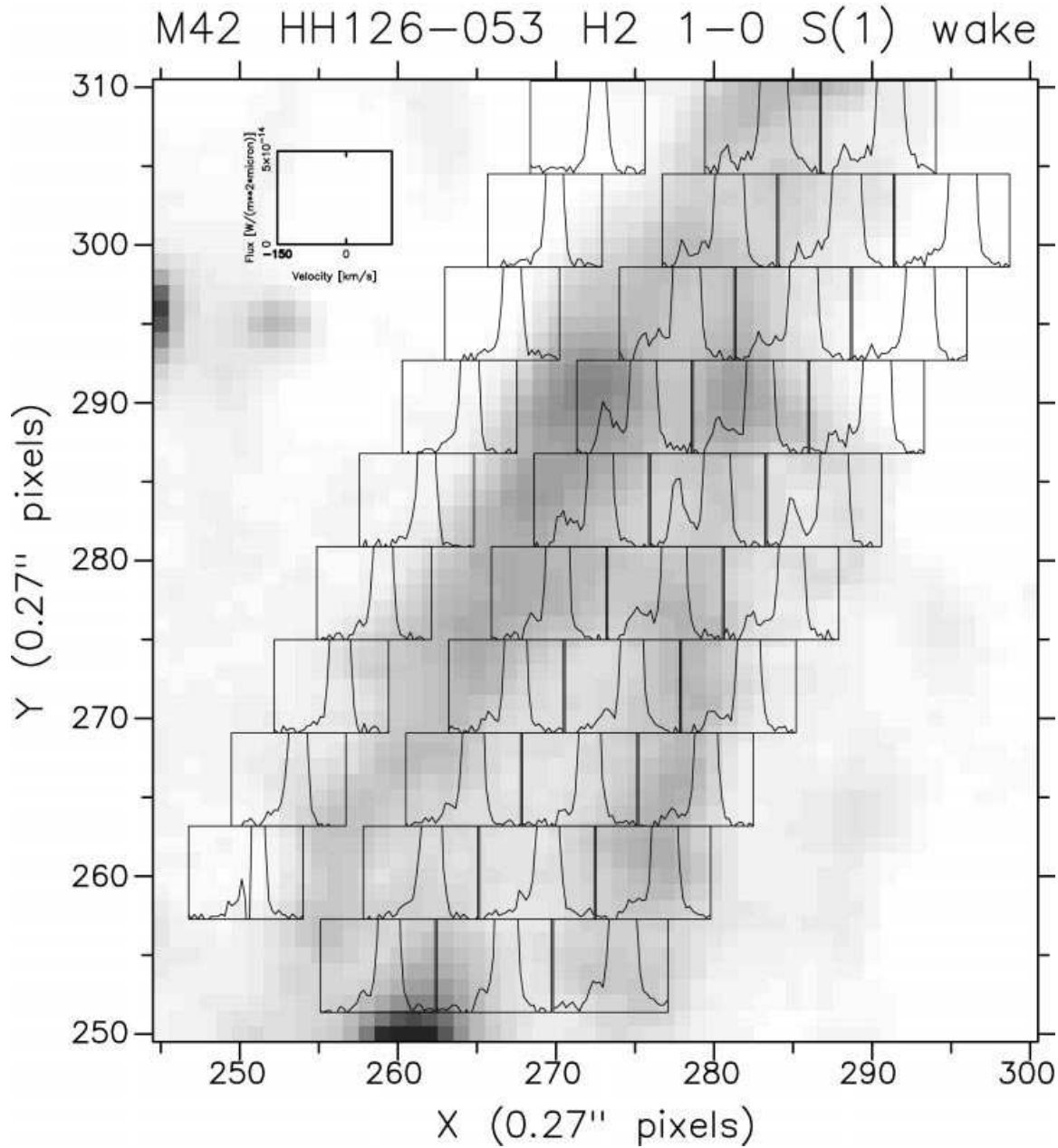

**Figure 15.** M42 HH126–053 background-subtracted $H_2$ 1–0 S(1) emission highlighting weak, high-velocity features including an artificial 'ghost' feature.

Burton (1993) images, we can measure the limb brightening and hence deduce an upper limit for the shell thickness of each of the two wakes studied to be $\lesssim 10^{16}$ cm. This limit on the post-shock cooling distance allows any steady-state shock model given recent observational limits on the magnetic field strength (Chrysostomou et al. 1994). Since the apparent bow shock structure is coherent over length-scales of many CGS4 pixels, we would expect to see a similar trend in peak velocity to that seen in [Fe II]. Then the resolved velocity widths of each expected $H_2$ peak depend on the shock type itself.

Bow shock models (e.g. Smith & Brand 1990c) predict broad, double-peaked emission profiles to be observed in the $H_2$ wakes, since a line of sight should intersect two distinct and oppositely directed shock fronts on either side of the bow shock wake. Peak separation will decrease from a maximum value near the head of the wake (exact position depending on orientation) to a minimum towards the tail, where the gas is only weakly shocked owing to the highly oblique angle between the shock front and the direction of propagation of the bow. Hence the profile width, proportional to the effective shock speed, will decrease for each component as one moves down the bow axis. A decrease in peak separation is also predicted in moving across from central axis positions to the outer limbs of the wake, perpendicular to the bow axis, as the direction of the shock impulse converges for each side of the wake. Singly peaked profiles are therefore not inconsistent with bow shock models for positions in the wake extremities, but the widths must be considerably smaller.

Carr (1993) was able successfully to interpret the multiply-peaked $H_2$ 1–0 S(1) 2.122-$\mu$m profile emission in the less clearly defined bow-shaped object HH7 (NGC 1333) as a bow shock.

**Table 4.** As Table 2 but for Gaussian line fits to any weak velocity component of background-subtracted H$_2$ 1–0 S(1) wake profiles in M42 HH126–053. 'U' indicates that the profile is unresolved. An instrumental ghost feature is not included (see text).

| Slit row | Slit B | | Slit C | | Slit D | | Slit E | |
|---|---|---|---|---|---|---|---|---|
| | $v_{\text{weak}}$ LSR /km s$^{-1}$ | intrinsic FWHM /km s$^{-1}$ | $v_{\text{weak}}$ LSR /km s$^{-1}$ | intrinsic FWHM /km s$^{-1}$ | $v_{\text{weak}}$ LSR /km s$^{-1}$ | intrinsic FWHM /km s$^{-1}$ | $v_{\text{weak}}$ LSR /km s$^{-1}$ | intrinsic FWHM /km s$^{-1}$ |
| 19 | – | – | −104.1±2.3 | 6.1±5.3 | −96.8±8.6 | 19.8±20.3 | – | – |
| 20 | – | – | −103.4±6.0 | 9.5±14.1 | −105.5±3.0 | U | – | – |
| 21 | – | – | −94.1±7.7 | 24.9±19.3 | −96.4±6.6 | 26.2±16.1 | – | – |
| 22 | −22.6±2.3 | U | −88.9±2.1 | 16.1±5.2 | −92.2±4.5 | 28.0±11.4 | −95.9±5.3 | 14.5±12.7 |
| 23 | −25.5±2.1 | 10.9±2.1 | −88.8±2.7 | 15.0±6.6 | −91.2±1.4 | 17.6±3.4 | −93.7±1.6 | 17.5±3.8 |
| 24 | – | – | – | – | −89.0±4.1 | 23.8±10.5 | −90.4±2.2 | 17.3±5.3 |
| 25 | – | – | – | – | – | – | −83.8±2.5 | 15.5±6.3 |
| 26 | −25.3±1.4 | 10.3±1.5 | – | – | – | – | −80.2±3.7 | 12.1±9.3 |
| 27 | – | – | −31.5±4.5 | U | −30.2±6.5 | U | −78.6±11.8 | U |
| 28 | – | – | −28.8±3.2 | U | – | – | – | – |

Furthermore, Fernandes & Brand (1995) successfully modelled the column densities of a range of H$_2$ transitions in the *K* band at the same two positions with a bow C-shock and localized fluorescence. Multiply peaked H$_2$ profiles were also resolved in the L1448 outflow by Davis & Smith (1996a), while some profile asymmetry was also observed in the massive DR 21 outflow (Davis & Smith 1996b).

We have observed H$_2$ 1–0 S(1) line profiles at positions along the brightest and most clearly resolved H$_2$ wakes in Orion (which are also the brightest and most clearly resolved in any source), but find them to be dominated by singly peaked, broad profiles centred within 10 km s$^{-1}$ of the peak background emission. This is extremely difficult to reconcile with any steady-state bow shock models. There are similarities to the singly peaked H$_2$ profiles observed in AS 353A/HH32 by Davis, Eisloffel & Smith (1996), although this is explained as a consequence of underlying bow asymmetry in that case. Our observations do show a gradual decrease in profile FWHM in moving down the wake, although this is less clear in the case of M42 HH120–114 owing to the presence of two wakes for some positions. Within individual wakes, however, this relation holds; but H$_2$ peak velocities that are only shifted by a few km s$^{-1}$ at most from the ambient velocity at positions where much larger shifts in [Fe II] peaks are seen are not consistent with this picture, although the largest such shifts ($\sim 8\,\text{km s}^{-1}$) are at positions at the head of the bows, which is expected.

Referring to Fig. 3, it is straightforward to show that the effective radial velocity ($v_{\text{rad}}$) observed for gas shocked by one side of a simple parabolic bow structure is given by the expression

$$v_{\text{rad}} = v_s \sin\psi \sin(\alpha - \psi), \tag{5}$$

where $v_s$ is the shock speed through the (stationary) ambient medium, $\psi$ is the angle between the tangent to the bow at that point and the bow axis, and $\alpha$ is the angle between the bow axis and the line of sight (orientation). For a line of sight that intersects regions on both sides of a wake, the radial velocity to the observer is oppositely directed for each component. Therefore two separate components are observed as long as the effective shock velocity is not so high as to dissociate H$_2$ in either component, as is the case for all positions in the Orion bullet wakes except those close to the tip associated with the [Fe II] 1.644-μm emission itself.

At positions near the head of the H$_2$ wake emission, e.g. row 20 of slit C in Table 3 for M42 HH126–053 (Fig. 14), one can use the bullet speed and orientation determined from the [Fe II] profiles together with the approximate bow position (hence $\psi$) in equation (5) to determine the peak velocity of the expected emission. In the simplest case of a J-shock this is equivalent to the shock velocity 'seen' by the gas. Fig. 18 shows kinematic model profiles calculated in this way over the relevant shock range sampled in a single CGS4 pixel (row 20) for each of the bullets M42 HH126–053 (slit C) and M42 HH120–114 (slit I). No dissociation speed limit has been imposed, so that the models merely indicate the wide peak separation and the relatively low emission strength near to the pre-shock velocity (set to zero), in contrast to the observed H$_2$ emission. While a peak velocity separation of this magnitude can be accommodated within the widest [Fe II] profiles at the tips of the bullet, they clearly cannot explain the singly peaked H$_2$ profiles given the relatively high velocity resolution (14 km s$^{-1}$ per velocity channel or 23 km s$^{-1}$ FWHM) of these observations compared with the profile widths. We should clearly resolve two separate peaks. In any planar shock, the peak H$_2$ emission velocity is shifted from that of the ambient cloud, since it is the downstream gas that is emitting. Although the peak emission for a C-shock occurs at velocities below the shock speed itself, it will still be at velocities significantly shifted from the pre-shock velocity of the gas, and always oppositely directed for each component sampled at positions in the near-plane-of-sky wakes observed here. Another possible mechanism is strong shocks followed by re-formation and then re-excitation, but this still does not explain the observed dynamics.

If the H$_2$ wakes are not single structures (e.g. if the ambient medium is highly clumped), we must break them down into smaller pieces. However, an explanation for the appearance of limb-brightened bow-shaped wakes is then required, especially given the success of the bow shock model in explaining the [Fe II] emission. In any case, each individual clump of shocked gas is still much bigger than the largest hydrodynamical drag lengths, and so high magnetic fields are required. It is only just possible to fit the observed profiles by theoretical C bow shock models even if the line of sight includes the entire bow shock and therefore the full, integrated velocity range. This can be seen by comparing with Smith & Brand (1990c, fig. 11) and assuming that the bow is oriented close to the plane of the sky and is moving at $v_s = 120\,\text{km s}^{-1}$, as determined from the [Fe II] profiles. Our observations, however, consist of small cuts through the bow, rather than the full range of shocked velocities sampled over the total bow surface. To fit a single C-shock absorber model (Smith et al. 1991) at these positions would imply an $H_2$ dissociation speed in excess of $\sim 80 \, \mathrm{km \, s^{-1}}$, with an implied magnetic field strength far in excess of observed estimates.

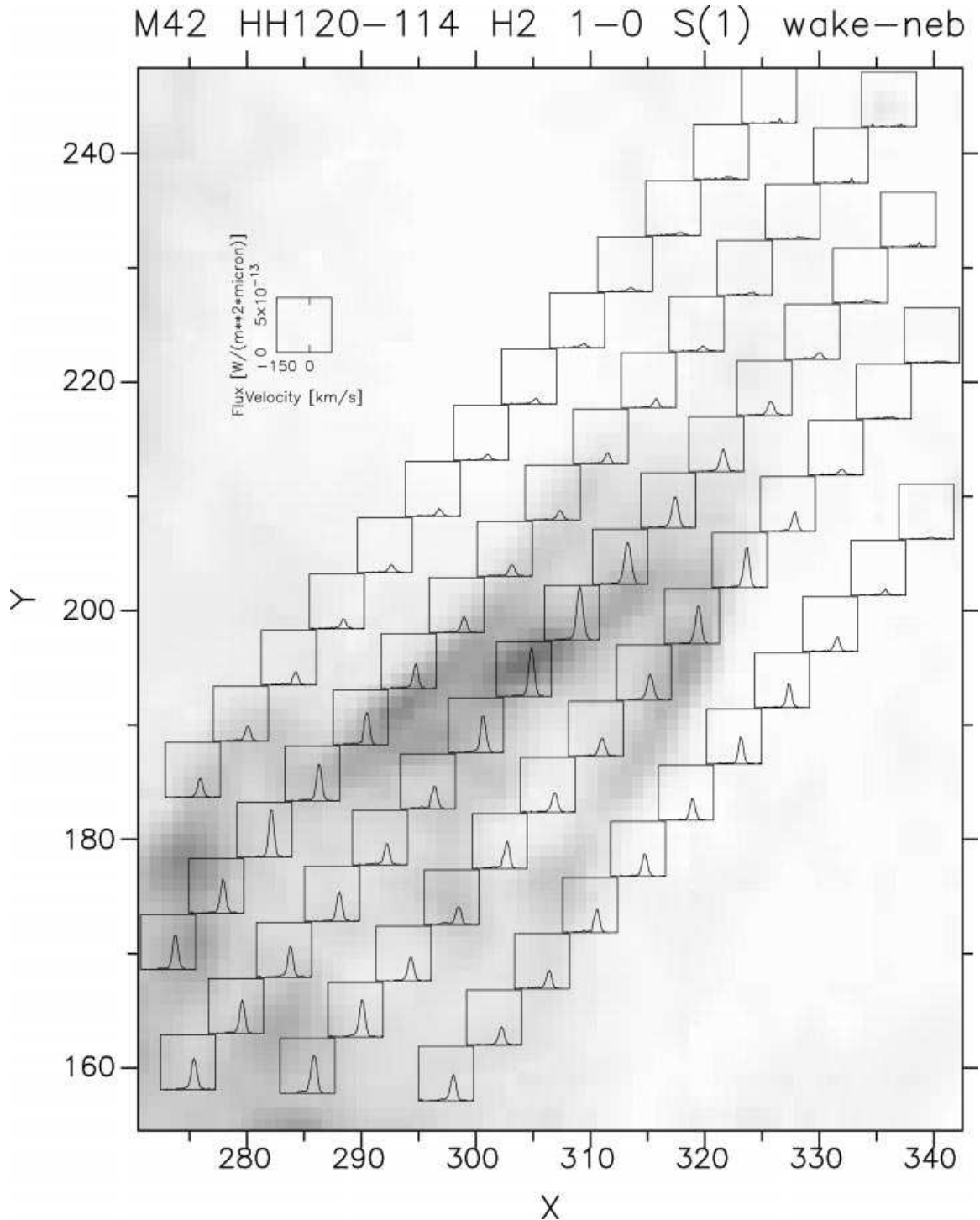

**Figure 16.** M42 HH120–114 background-subtracted $H_2$ 1–0 S(1) velocity profiles associated with the bullet.

Norris (1984) estimated that a 3-mG magnetic field strength could explain the apparent Zeeman splitting observed in OH masers in Orion–KL. Later polarization measurements by Chrysostomou et al. (1994) allow field strengths as high as $\sim 10 \, \mathrm{mG}$, but the measured value is sensitive to estimates of the density and turbulent velocity. The value quoted is determined by assuming $n \sim 10^6 \, \mathrm{cm^{-3}}$, which is in agreement with shock modelling (Brand et al. 1988) and detection of $J = 3$–2 transitions of HCN and $HCO^+$ (White et al. 1986). The turbulent velocity is estimated to be of the order of $\sim 1 \, \mathrm{km \, s^{-1}}$ from the FWHM of the Doppler-broadened lines of CS $J = 2$–1 transitions (Mundy et al.

**Table 5.** Gaussian line fits to the main velocity component of background-subtracted $H_2$ 1–0 S(1) wake profiles in M42 HH120–114. Velocities are calibrated relative to the dynamical Local Standard of Rest.

| Slit row | Slit H | | Slit I | | Slit J | | Slit K | | Slit L | |
|---|---|---|---|---|---|---|---|---|---|---|
| | $v_{peak}$ /km s$^{-1}$ | intrinsic FWHM /km s$^{-1}$ | $v_{peak}$ /km s$^{-1}$ | intrinsic FWHM /km s$^{-1}$ | $v_{peak}$ /km s$^{-1}$ | intrinsic FWHM /km s$^{-1}$ | $v_{peak}$ /km s$^{-1}$ | intrinsic FWHM /km s$^{-1}$ | $v_{peak}$ /km s$^{-1}$ | intrinsic FWHM /km s$^{-1}$ |
| 19 | 5.3±2.3 | 44.5±5.4 | 5.4±2.1 | 36.4±5.0 | 7.5±0.9 | 25.5±2.2 | −4.4±4.6 | 69.3±10.9 | −7.6±6.6 | 91.5±15.6 |
| 20 | 1.4±1.6 | 35.0±3.8 | 3.0±1.0 | 26.1±2.4 | 4.9±0.3 | 22.6±0.9 | 1.7±1.0 | 26.0±2.3 | −1.7±4.6 | 80.1±10.9 |
| 21 | 4.0±1.5 | 26.1±3.5 | 7.9±0.5 | 12.9±1.3 | 6.9±0.2 | 23.1±0.5 | 6.1±0.3 | 15.9±0.7 | 6.8±1.1 | 16.7±2.5 |
| 22 | 4.1±0.8 | 18.4±2.0 | 6.7±0.5 | 14.1±1.1 | 6.7±0.2 | 25.1±0.5 | 6.0±0.2 | 17.2±0.6 | 6.6±0.4 | 15.7±0.9 |
| 23 | 4.8±0.9 | 20.9±2.1 | 8.1±0.5 | 19.6±1.1 | 9.8±0.1 | 27.3±0.4 | 5.2±0.2 | 19.9±0.5 | 6.6±0.2 | 12.6±0.5 |
| 24 | 4.5±0.6 | 20.4±1.5 | 8.3±0.4 | 21.4±0.9 | 10.5±0.1 | 24.1±0.4 | 4.3±0.2 | 23.3±0.5 | 5.9±0.2 | 10.4±0.5 |
| 25 | 4.9±0.6 | 19.1±1.4 | 9.2±0.3 | 16.3±0.8 | 8.5±0.1 | 18.3±0.4 | 4.1±0.3 | 21.6±0.7 | 4.4±0.2 | 13.1±0.5 |
| 26 | 5.3±0.5 | 16.1±1.3 | 7.4±0.2 | 13.3±0.6 | 7.6±0.2 | 16.6±0.5 | 6.9±0.2 | 19.4±0.7 | 5.5±0.2 | 16.8±0.6 |
| 27 | 5.8±0.5 | 12.4±1.2 | 4.0±0.2 | 13.5±0.5 | 4.6±0.4 | 19.0±0.9 | 7.6±0.3 | 18.8±0.8 | 6.0±0.4 | 15.8±1.0 |
| 28 | 7.0±0.3 | 15.6±0.5 | 5.3±0.1 | 14.7±0.4 | 6.3±0.2 | 15.2±0.7 | 6.3±0.3 | 17.0±0.7 | 6.6±0.3 | 14.9±0.7 |

1988). However, if the velocity is as high as 3 km s$^{-1}$, this combines with a maximum uncertainty of up to a factor of 10 in the density to give a range of ∼ 3–95 mG for the strength of the magnetic field. Ionization fractions in excess of $10^{-5}$ to $10^{-4}$ in the pre-shock gas would tie the neutral fluid to the ionized fluid, effectively freezing the magnetic field. This would eliminate the magnetic precursor and make the shock front J-type (Smith 1994).

The detection of weak but extremely high-velocity (peak velocity ≲ −105 km s$^{-1}$) $H_2$ emission features at positions close to or coincident with the [Fe II] bullet emission in both wakes is inexplicable with any steady-state molecular shock models, unless we are not resolving multiple shock fronts along the line of sight. Micono et al. (1998) suggest a Mach disc origin for a compact $H_2$ knot associated with the head of the HH46/47 counter-bow shock. However, Mach discs represent the strongest shocks in a flow where the surroundings are stationary, and so would destroy $H_2$. Only if the shocks were ploughing into previously nearly isokinetic gas would the Mach disc be a sufficiently weak shock, but then the velocity ought to show in the overall pattern as a bulk velocity, which is not observed. We note that the peak velocity of the individual features identified in M42 HH126–053 clearly moves closer to the main emission peak in moving down the wakes, and so appears to be associated with the bullets. The presence of the instrumental ghost feature in between the strong and weak components, together with the relatively poor signal-to-noise ratio, however, prevents a more detailed analysis. When resolved, the FWHM of Gaussian line fits to these components shows no clearly discernible trend with position, but never exceeds the FWHM of the strong, zero-velocity centred emission.

In the absence of the expected double-peaked $H_2$ profiles, we examine the possibility that the fingers of $H_2$ may contain dust at a density that is high enough to extinguish $H_2$ emission completely from the far side of the wake (relatively redshifted emission). To determine the implied gas density required for this, we set the optical depth, $\tau$, given by

$$\tau = \int_{\text{pathlength}} \rho \alpha_{\text{ext}} dx, \qquad (6)$$

to unity. Here, $\rho$ is the dust mass density (g cm$^{-3}$), $\alpha_{\text{ext}}$ is the dust extinction cross-section per unit mass at 2.122 $\mu$m (cm$^2$ g$^{-1}$) and $x$ is the distance along the line of sight. The integral is from one face of the shock structure to the other. Examining the Allen & Burton (1993) images, the wake structures are of the order of $10^{17}$ cm across. Combining this with a value for the dust opacity of $\alpha_{\text{ext}} = 3.25 \times 10^3$ cm$^2$ g$^{-1}$ at 2.512 $\mu$m (Ossenkopf & Henning 1994) gives an implied mass density of $3.08 \times 10^{-21}$ g cm$^{-3}$. Using the typical interstellar medium value $\rho_{\text{dust}} \simeq 0.01 \rho_{\text{gas}}$ with the observed extinction, and knowing that $m_{H_2} = 3.3 \times 10^{-24}$ g, therefore implies an $H_2$ gas number density of ∼ $10^5$ cm$^{-3}$, which is similar to the inferred densities in the OMC-1 region. Clearly this possibility will require further investigation.

### 5.3 Alternative mechanisms

Tedds et al. (1995) suggested that the effects of instabilities might be important for the radiative cooling observed in outflows such as Orion, and indeed more recent imaging and velocity-resolved spectroscopy in AS 353A/HH32 (Davis et al. 1996), the DR 21 outflow (Davis & Smith 1996b) and the supernova remnant IC 443 (Richter 1995) also show singly peaked profiles at positions within apparently bow-shaped $H_2$ 1–0 S(1) morphologies.

Given that the shock fronts will be oblique along a bow, the

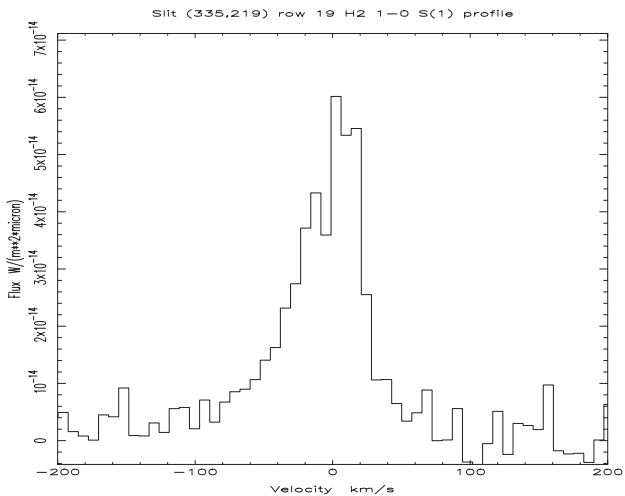

**Figure 17.** Individual $H_2$ 1–0 S(1) velocity profile at slit K row 19, upstream of the prominent M42 HH120–114 wake emission but coincident with strong [Fe II] emission. The relatively narrow and weak background emission has not been subtracted in this case, for higher signal-to-noise ratio. Note the very large total velocity range near zero intensity and the similarity to the corresponding [Fe II] profile in Fig. 8.

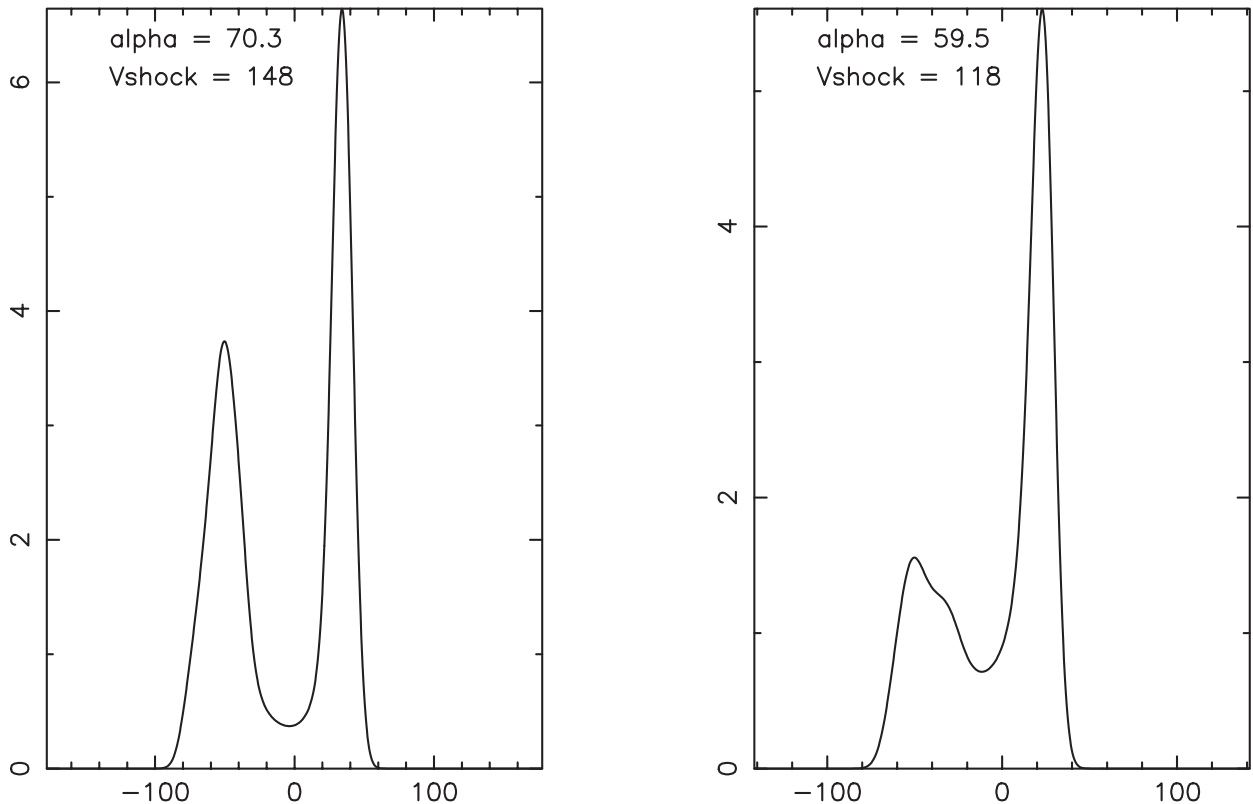

**Figure 18.** Kinematic model predictions of possible shocked $H_2$ 1–0 S(1) 2.122-$\mu$m line profiles in the following pixel positions. Left: M42 HH126–053 slit C, row 20; right: M42 HH120–114 slit I, row 20. The model assumes that the post-shock relative velocity is negligible and that the shocks move into a stationary medium. No dissociation speed limit has been imposed on the shocked $H_2$ and profiles are not smoothed by instrumental broadening.

Wardle instability in the case of C-shocks was investigated by Wardle and Draine (1987) & Wardle (1991a,b). However, recent numerical simulations indicate that such an instability, although becoming significant over long time-scales, will not alter the $H_2$ radiation emitted on relatively short time-scales at the shock fronts themselves (Stone 1997; Neufeld & Stone 1997; MacLow & Smith 1997). Turbulent boundary layers were investigated by Dyson et al. (1995), but in this case one would still expect separate emission peaks to be resolved at each of the effective shock velocities seen by each side of a spectroscopic cut through a plane-of-sky-oriented bow shock.

Stone, Xu & Mundy (1995) were also able to reproduce the large-scale OMC-1 morphology in numerical simulations of Rayleigh–Taylor instabilities produced by a time-varying wind as an alternative launch mechanism for the fast-moving bullets. This is supported by spectroscopic imaging of the central OMC-1 region by McCaughrean & MacLow (1997), but an alternative 'thin layer instability' in rapidly cooling radiative shock zones is suggested by Schild, Miller & Tennyson (1997). Importantly, the effects of supersonic turbulence and turbulent decay time-scales (owing to dissipation of energy in shock waves) are as yet little understood, and may be expected to alter significantly the excited $H_2$ level populations compared with steady-state models. We go on to measure the $H_2$ excitation spectrum at the same positions within the Orion bullets in a forthcoming paper [Tedds et al., in preparation (Paper II)].

### 5.4 The nature of background contribution(s) to profiles

In front of the shocked line emission from OMC-1 there is a zone of fluorescent $H_2$ line emission, resulting from the UV excitation of the molecular cloud by the Trapezium stars. While in the line of sight to OMC-1 it is clear that shocked/thermal emission dominates, it is important to estimate the contribution of fluorescent emission, especially as its profile should be unresolved by our observations, being intrinsically narrow compared with collisionally broadened $H_2$ emission. Large-scale diffuse $H_2$ mapping by Luhman et al. (1994) over the entire Orion A cloud in the 6–4 Q(1) line at 1.601 $\mu$m and the 1–0 S(1) line shows that UV fluorescence dominates the emission in the outer parts of the cloud and accounts for $\sim$ 98–99 per cent of the global $H_2$ emission. Burton & Puxley (1990) showed that shocks alone would produce relatively more 1–0 S(1) emission, contributing $\sim$ 7 per cent of the total $H_2$ line flux. The line profiles of the fluorescent emission from the Orion Bar are unresolved spectrally with FWHM $\leq 17\,\text{km}\,\text{s}^{-1}$ (Burton et al. 1990) and are centred at the ambient cloud velocity. This is in clear contrast to the $140\,\text{km}\,\text{s}^{-1}$ wide profile with a broad blue wing at OMC-1 (Brand et al. 1989b). Hence, if fluorescence were dominating the emission in any of the profiles that we measured, we would expect to see a significantly enhanced, narrow component to the profile at the rest velocity, compared with the shock profiles.

Following the analysis of Burton & Puxley (1990), the fluorescent contribution to the 1–0 S(1) line at the peak of OMC-1 is only about 1 per cent, but moving to the edges of the strong emission regions it rises to about 10 per cent. In the region containing the observed bullet wakes, the strength of the fluorescence has not been measured in detail. In both tracers measured here, the profile of the background emission pervading the observed bullet region is well fitted by a single Gaussian profile centred within $3\,km\,s^{-1}$ of zero velocity relative to the dynamical Local Standard of Rest. However, the $H_2$ profile is not unresolved, as would be the case if it were excited by fluoresence alone, but is significantly broadened by $26.0 \pm 2.5\,km\,s^{-1}$ at FWHM. Therefore the background component can perhaps be modelled as a combination of an unresolved fluorescent component together with a uniform, collisionally broadened component. If the broadening is interpreted as purely thermal, this corresponds to an rms velocity $v_{rms} = 16.8\,km\,s^{-1}$ and a uniform excitation temperature of $\simeq 34\,000\,K$!

It has been suggested that the broad component may be the result of an expanding bubble (Chrysostomou et al. 1997), centred near the BN–IRc2 complex, which pushes a shock front through the molecular gas as it traverses the entire observed region. Norris (1984) refers to an isotropic source of weak OH maser emission about IRc2, analogous to the 'low-velocity' $H_2O$ masers reported by Genzel et al. (1981), in addition to clusters of stronger features at distinct positions. If the emission were due to cooling upon passage of a uniform, plane J- or C-shock front through ambient molecular gas, we would expect the peak velocity of the resultant profiles to vary with position in the region as the line of sight intersects components deflected at differing angles and varying apparent shock velocity. We would also expect to see two different components, separated in velocity, along a given line of sight intersecting two different sides of the bubble. This appears to rule out this explanation, as, even though the underlying cloud geometry is unknown, it is difficult to model the emission resulting in such a spatially uniform, singly peaked profile.

For individual profiles, the broadening may be explained by a single, highly magnetized planar C-shock. We are then confronted with the problem of how to excite the corresponding [Fe II] emission in a C-shock over the same range of velocities. The constancy of line shape, width and peak velocity, however, can only then be explained if the shock is seen face-on only over the entire region, or perhaps if we observe a full range of unresolved shock fronts along each line of sight as a uniform wind impacts highly clumped molecular gas. Alternatively, we may again be observing the consequences of supersonic turbulence or instabilities. We go on to measure the $H_2$ excitation spectrum of this background component and the bullet wakes in a forthcoming paper (Paper II).

## 6 CONCLUSIONS

We have demonstrated that integrated [Fe II] line profiles in the Orion bullets M42 HH126–053 and M42 HH120–114 are consistent with theoretical bow shock predictions. We have identified a uniform, broad background component pervading the region in both [Fe II] and $H_2$ which is inconsistent with a fluorescent component arising from the ionizing radiation of the Trapezium stars alone. A collisionally broadened background component of unidentified origin is measured with an average FWHM of $26 \pm 2.5\,km\,s^{-1}$ in the $H_2$ 1–0 S(1) line and a peak velocity of $2.5 \pm 0.5\,km\,s^{-1}$, close to the local ambient rest velocity.

The extended $H_2$ bullet wakes have allowed us to dissect individual molecular bow shock structures, but the broad (FWHM $\leq 26\,km\,s^{-1}$), singly peaked $H_2$ 1–0 S(1) profiles (where $v_{peak}$ varies by only a few $km\,s^{-1}$ from the background) observed in the two most clearly resolved, plane-of-sky-oriented wakes challenge our present understanding. It is very difficult to reconcile *any* steady-state bow shock model with these observations in Orion. To fit a single C-shock absorber model to individual profiles implies a magnetic field strength far in excess of observed estimates, and is not consistent with the bow-shaped wake morphology.

Alternatively, we may still not be resolving multiple shock fronts along the line of sight. For example, multiple overlapping bullet wakes could give rise to merged sets of doubly peaked profiles, resulting in approximately Gaussian-shaped profiles. However, given the appearance of single bow-shaped wakes at many observed positions, the accuracy of pure Gaussian line fits, the velocity resolution of our observations and the fact that we see this phenomenon in *two* different wakes, this explanation requires very tight constraints on the numbers of unresolved clumps within the small ($\sim$ 1-arcsec pixel) beamsize of these observations.

If we cannot fit the profiles in Orion with steady-state molecular shocks, it may be necessary to model the effects of instabilities and turbulence. This will have important consequences. Not only will line profiles be broadened, but level populations of shocked species will be altered, and hence so will the observed column densities over a range of transitions. New observations of $H_2$ column densities in these bullet wakes (Paper II) will address this.

## ACKNOWLEDGMENTS

UKIRT is operated by the Joint Astronomy Centre, Hawaii, on behalf of the UK Particle Physics and Astronomy Research Council (PPARC). We thank all the UKIRT staff for their excellent support and assistance, and Horst Meyerdierks for developing the SPECDRE SPECGRID routine in Starlink used to display this work. Thanks go to Amadeu Fernandes, Antonio Chrysostomou, Henry Buckley and Tom Geballe for useful discussions. We thank the referee Chris Davis for useful suggestions and comments. JAT acknowledges a research studentship award from PPARC and a University Fellowship from the University of Leeds. PWJLB acknowledges support from the Kerr–Fry Bequest, the Anglo-Australian Observatory, the RCFTA Fund of the University of Sydney and the University of New South Wales while this work was in progress.